\documentclass[nonacm,sigplan,table,balance]{acmart}

\pdfoutput=1

\AtBeginDocument{%
  \providecommand\BibTeX{{%
    \normalfont B\kern-0.5em{\scshape i\kern-0.25em b}\kern-0.8em\TeX}}}

\usepackage[switch]{lineno}
\usepackage{adjustbox}
\usepackage{algorithm}
\usepackage{algpseudocode}
\usepackage{amsfonts}
\usepackage{amsmath}
\usepackage{amsthm}
\usepackage{appendix}
\usepackage{booktabs}
\usepackage{colortbl}
\usepackage[inline]{enumitem}
\usepackage{ifthen}
\usepackage{makecell}
\usepackage{multirow}
\usepackage{pifont}
\usepackage[binary-units]{siunitx}
\usepackage{subcaption}
\usepackage{textcomp}
\usepackage{tikz}
\usepackage{xspace}

\usepackage[capitalize,sort&compress]{cleveref}

\newlist{myinlinelist}{enumerate*}{1}
\setlist[myinlinelist]{label=(\roman*)}

\newtheorem{prop}{Property}
\crefname{prop}{Property}{Properties}
\Crefname{prop}{Property}{Properties}

\DeclareSIUnit{\cpuyear}{CPU\textrm{-}years}
\DeclareSIUnit{\lines}{lines}

\graphicspath{{./}{images/}}

\usetikzlibrary{shapes,arrows,trees,matrix,positioning,decorations.pathreplacing}
\tikzset{
    brace/.style = {decorate, decoration={brace, amplitude=10pt} },
    mbrace/.style = { decorate, decoration={brace, amplitude=10pt, mirror} },
    label/.style = { black, midway, scale=0.5, align=center },
    toplabel/.style = { label, above=.5em, anchor=south, font=\large },
    leftlabel/.style = { label,rotate=-90,left=.5em,anchor=north, font=\large },
    bottomlabel/.style = {label, below=1em, anchor=north },
    sumlabel/.style = {font=\large,below=1ex}
}

\newcommand{\mygrid}[3][0]{ 
    \foreach \row in {1,...,#2} 
    \foreach \col in {1,...,#3} 
    {
        \draw[thick] (\col,\row) +(-0.5,-0.5) rectangle ++(+0.5,+0.5);
        \ifthenelse{\col=1}{ 
            \node[] at (\col-1,\row) {\large $s_{\row}$};
        }{} 
        \ifthenelse{\row=#2}{ 
            \node[] at (\col,\row+1) {\large $e_{\col}$};
        }{} 
    }
}

\newcommand{\bitlydata}{\url{https://bit.ly/2JJfffY}\xspace}
\newcommand{\bitlytools}{\url{https://bit.ly/2WZVynP}\xspace}

\newcommand{\tool}{MoonLight\xspace}
\newcommand{\bvtool}{MoonBeam\xspace}
\newcommand{\cmin}{\texttt{afl-cmin}\xspace}

\newcommand{\showmap}{\texttt{afl-showmap}\xspace}
\newcommand{\minset}{\textsc{Minset}\xspace}
\newcommand{\uminset}{\textsc{Unweighted Minset}\xspace}
\newcommand{\coverset}{\ensuremath{\mathcal{C}}\xspace}

\newcommand{\numdistills}{five\xspace}
\newcommand{\numftsprogs}{\num{10}\xspace}
\newcommand{\numrealprogs}{six\xspace}
\newcommand{\numtotalprogs}{16\xspace}
\newcommand{\realnumfiletypes}{seven\xspace} 
\newcommand{\numfiletypes}{13\xspace} 
\newcommand{\numbugs}{\num{33}\xspace}
\newcommand{\numnewbugs}{nine\xspace}
\newcommand{\numcves}{seven\xspace}
\newcommand{\numcollectedseeds}{\num{2823412}\xspace}
\newcommand{\numcandidateseeds}{\num{944375}\xspace}

\newcommand{\numcpuyears}{\SI{34}{\cpuyear}\xspace}

\colorlet{rowgray}{gray!15}
\newcommand{\bcell}{\cellcolor{hlblue}}
\newcommand{\gcell}{\cellcolor{hlgreen}}
\newcommand{\ycell}{\cellcolor{hlyellow}}

\newcommand{\bugdesc}[1]{\multicolumn{8}{l}{\textit{#1}}}

\definecolor{hlyellow}{HTML}{FFFFC7}
\definecolor{hlgreen}{HTML}{C7FFC9}
\definecolor{hlblue}{HTML}{C7EEFF}

\begin{document}

\title{Corpus Distillation for Effective Fuzzing}
\subtitle{A Comparative Evaluation}

\author{Adrian Herrera}
\authornote{Both authors contributed equally to this research.}
\email{adrian.herrera@anu.edu.au}
\affiliation{%
  \institution{Defence Science and Technology Group \&
    Australian National University}
}

\author{Hendra Gunadi}
\authornotemark[1]
\email{hendra.gunadi@anu.edu.au}
\affiliation{%
  \institution{Australian National University}
}

\author{Liam Hayes}
\author{Shane Magrath}
\affiliation{%
  \institution{Defence Science and Technology Group}
}

\author{Felix Friedlander}
\author{Maggi Sebastian}
\affiliation{%
  \institution{Australian National University}
}

\author{Michael Norrish}
\affiliation{%
  \institution{Data61 \& Australian National University}
}

\author{Antony L.\ Hosking}
\affiliation{%
  \institution{Australian National University \& Data61}
}

\renewcommand{\shortauthors}{Herrera and Gunadi, et al.}

\begin{abstract}
Mutation-based fuzzing typically uses an initial set of non-crashing
seed inputs (a corpus) from which to generate new inputs by
mutation. A given corpus of potential seeds will often contain
thousands of similar inputs. This lack of diversity can lead to wasted
fuzzing effort, as the fuzzer will exhaustively explore mutation from
all available seeds. To address this, fuzzers such as the well-known
American Fuzzy Lop (AFL) come with \emph{distillation} tools (e.g.,
\cmin) that select seeds as the smallest subset of a given corpus that
triggers the same range of instrumentation data points as the full
corpus. Common practice suggests that minimizing both the
\emph{number} and \emph{cumulative size} of the seeds may lead to more
efficient fuzzing, which we explore systematically.

We present results of over \numcpuyears of fuzzing with \numdistills
distillation alternatives to understand the impact of distillation on
\emph{finding bugs in real-world software}.  We evaluate a number of
existing techniques---including \cmin and \minset---and also present
\emph{\tool}: a freely available, configurable, state-of-the-art,
open-source, distillation tool.

Our experimental evaluation compares the effectiveness of distillation
approaches, targeting the Google Fuzzer Test Suite and a diverse set
of \numrealprogs real-world libraries and programs, covering
\numfiletypes different input file formats across \numtotalprogs
programs. Our results show that distillation is a necessary precursor
to \emph{any} fuzzing campaign when starting with a large initial
corpus. We compare the effectiveness of alternative distillation
approaches. Notably, our experiments reveal that state-of-the-art
distillation tools (such as \tool and \minset) do not exclusively find
all of the \numbugs bugs (in the real-world targets) exposed by our
combined campaign: each technique appears to have its own strengths.
We find (and report) new bugs with \tool that are not found by
\minset, and \textit{vice versa}.  Moreover, \cmin fails to reveal
many of these bugs.  Of the \numbugs bugs revealed in our campaign,
\numcves new bugs have received CVEs.



\end{abstract}



\maketitle

\section{Introduction}\label{sec:Intro}

Fuzzing is a dynamic analysis technique for finding bugs and
vulnerabilities in software, aiming to trigger crashes in a target
program by subjecting it to a large number of (possibly malformed)
inputs.  \emph{Mutation-based} fuzzing typically uses an initial set
of valid seed inputs from which to generate new seeds by random
mutation. A given corpus of potential seeds will often contain
thousands of inputs that generate similar behavior in the target,
which can lead to wasted fuzzing effort in exhaustive mutation from
all available seeds.

Due to their simplicity and ease of use, mutation-based fuzzers such as
AFL~\cite{AFL:2015}, honggfuzz~\cite{honggfuzz:2016}, and
libFuzzer~\cite{libFuzzer:2016} are widely deployed in industry, where they
have been highly successful in uncovering thousands of bugs across a large
number of popular programs~\cite{OSSFuzz:2016, OSSFuzz:2017}.  This success has
prompted much research into improving various aspects of the fuzzing process,
including mutation strategies~\cite{Steelix:2017}, seed selection
policies~\cite{Collafl:2018}, and path-exploration
algorithms~\cite{Qsym:2018}.

In addition, researchers often cite the importance of high-quality
input seeds and their impact on fuzzer performance~\cite{Wang2017,
  Hicks:2018, Rebert:2014, Pailoor2018}.  However, relatively few
studies address the problem of \emph{optimal design and construction
  of corpora} for mutation-based fuzzers
\cite{Rebert:2014,Pailoor2018}.  Intuitively, there are several
properties one might desire of the collection of seeds that form the
initial corpus:

\begin{prop}[Maximize coverage of target behaviors]
  \label{corpus-prop:coverage}
  Seeds in the corpus should generate a broad range of observable behaviors in
  the target.  Fuzzers typically approximate this with code coverage, so the
  seeds should collectively exercise as much code as possible.  Lack of
  coverage diversity inhibits exploration of behavior during fuzzing.
\end{prop}

\begin{prop}[Eliminate redundancy in seed behavior]
  \label{corpus-prop:redundant-seeds}
  Candidate seeds that are behaviorally similar to one another (following from
  \cref{corpus-prop:coverage}: that produce the same code coverage) should be
  represented in the corpus by a single seed.  Fuzzing multiple seeds with the
  same behavior is wasteful~\cite{Rebert:2014}.
\end{prop}

\begin{prop}[Minimize the total size of the corpus]
  \label{corpus-prop:min-corpus-size}
  This reduces storage costs and results in a significant reduction of the mutation search space.
\end{prop}

\begin{prop}[Minimize the sizes of the seeds]
  \label{corpus-prop:min-seed-size}
  Contention in the storage system should be avoided where possible.
  Fuzzers are highly I/O bound, so smaller seed files should be preferred to
  reduce I/O requests to the storage system~\cite{Min:2016:FxMark, Xu:2017}.
  In turn, this will shorten the execution time of each iteration,
  achieving more coverage in any fixed amount of time.
\end{prop}

Under these assumptions, simply gathering as many input files as possible is
not an optimal approach for constructing a fuzzing corpus (due to
\cref{corpus-prop:redundant-seeds,corpus-prop:min-corpus-size,corpus-prop:min-seed-size}
above).  Conversely, these assumptions also suggest that beginning with the
``empty corpus'' (e.g., consisting of one zero-length file or a valid seed
having minimal coverage) may be less than ideal (due to
\cref{corpus-prop:coverage}).

The following natural questions arise:
\begin{myinlinelist}
\item How do we best select seeds for a fuzzing corpus?
\item If we assume \crefrange{corpus-prop:coverage}{corpus-prop:min-seed-size}
  above, how should they be weighted with respect to each other?
\item Having generated answers to our first two questions, does the resulting
  approach produce corpora that in turn produce better results (more bugs for
  the same amount of fuzzing time) than alternative, state-of-the-art
  approaches?
\end{myinlinelist}

We call the process of seed selection \emph{corpus
  distillation}.\footnote{Distillation might also be referred to as
  \emph{reduction} or \emph{minimization}.  We choose to use the same
  language as \citet{Pailoor2018}, and avoid the term \emph{reduction}
  since it is also used in the crash triage process to reduce crash
  exemplars to a minimum size~\cite{Zeller:2002, Groce:2014}.}  We
assume that we already have a large candidate corpus, in the form of
inputs already gathered, and so we can also explore distillation
strategies that range from throwing away all the seeds entirely
through to keeping all the candidate seeds.  In this way, we can test
our assumptions above.

\subsection{Contributions}
We comprehensively evaluate a number of corpus distillation techniques developed
and used by both academia and industry:

\paragraph{\tool}
We design and implement a new corpus distillation
tool---\emph{\tool}---which represents distillation as a (weighted)
minimum set cover problem (WMSCP) and efficiently computes a solution
using a dynamic programming approach.  In so doing, we extend the
\minset approach \cite{Rebert:2014} to develop a new theory for corpus
distillation as the foundation for \tool (\cref{sec:techniques}).

\paragraph{Comprehensive Evaluation}
We perform comprehensive and rigorous evaluation of \numdistills
corpus distillation techniques---including the widely-used \cmin, the
state-of-the-art \minset, and our \tool---comparing resulting corpus
sizes and bug-finding ability.  In particular, we evaluate bug-finding
ability by means of extensive fuzzing campaigns over a diverse set of
target programs, including the Google Fuzzer Test Suite and
\numrealprogs popular open-source libraries.  We also evaluate the
extreme points of the distillation spectrum, comprising the full
(undistilled) corpus and the empty corpus (\cref{sec:Evaluation}).

\paragraph{Crash Triage}
Because the ultimate aim of fuzzing is to uncover bugs in software, we
have triaged all crashes, and find that no one distillation technique
finds all of the bugs discovered during our fuzzing campaigns.  Both
\tool and \minset appear to have their own strengths, while also
producing smaller corpora than the widely-used \cmin.  Many of the
\numbugs bugs found in our real-world target set were known (and
security-interesting; see Table~\ref{tab:triage-results} for details),
but \numcves were both security-interesting and previously
undiscovered.  For these \numcves, we have logged bug reports and
received CVEs.


\section{Background}\label{sec:LitReview}



Fuzzing has become a popular technique for automatically finding bugs
and vulnerabilities in software.  This popularity can be attributed to
its simplicity and success in finding bugs in ``real-world''
software~\cite{OSSFuzz:2017, AFL:2015, honggfuzz:2016,
  libFuzzer:2016}.  At a high level, fuzzing involves the generation
of large numbers of test-cases that are fed into the target program to
induce a crash.  The target is monitored so that crash-inducing
test-cases can be identified and saved for further analysis/triage
after the fuzzing campaign has ended.

How a fuzzer generates test-cases depends on whether it is
\emph{generation}-based or \emph{mutation}-based.  Generation-based
fuzzers (e.g., QuickFuzz~\cite{QuickFuzz:2017},
Dharma~\cite{Dharma:2015}, and
CodeAlchemist~\cite{CodeAlchemist:2019}) require a specification/model
of the input format.  They use this specification to synthesize
test-cases.  In contrast, mutation-based fuzzers (e.g.,
AFL~\cite{AFL:2015}, honggfuzz~\cite{honggfuzz:2016}, and
libFuzzer~\cite{libFuzzer:2016}) require an initial corpus of seed
inputs (e.g., files, network packets, and environment variables) to
bootstrap test-case generation.  New test-cases are then generated by
mutating inputs in this initial corpus.

Perhaps the most popular mutation-based fuzzer is American Fuzzy Lop
(AFL)~\cite{AFL:2015}.  AFL is a \emph{greybox} fuzzer, meaning that it uses
light-weight instrumentation to gather \emph{code coverage} information during
the fuzzing process.  This code coverage information acts as an approximation
of program behavior.  AFL instruments edge transitions between basic blocks
and uses this information as code coverage.  By feeding the code coverage
information back into the test-case generation algorithm, the fuzzer is able
to explore new executions (and hence behaviors) in the target.  In addition to
the core fuzzer, AFL also provides a corpus distillation tool: \cmin
(discussed further in \cref{sec:techniques}).

\subsection{Formalizing the Distillation Problem}
\label{sec:FormalCorpusDistillation}

Our work focuses on solving the problem of optimal design and construction of
corpora for mutation-based fuzzers.  To solve this problem, the primary
question we need to answer is that posed by \citet{Rebert:2014}:
\begin{quote}
  Given a large collection of inputs for a particular target (the
  \emph{collection corpus}), how do we select a subset of inputs that
  will form the initial \emph{fuzzing corpus}?
\end{quote}
We refer to the process of selecting this subset of inputs as
\emph{distillation}.  In particular, we are most interested in distillation
that leads to \emph{more efficient fuzzing}.  As the ultimate aim of fuzzing
is to uncover bugs in software, this means producing a \emph{higher bug yield}
than if we had simply used the collection corpus as the fuzzing corpus.  This
is because most seeds in a collection corpus are behaviorally very similar to
each other.
Therefore it is important to distill the possibly very large collection corpus into a much smaller fuzzing corpus, which is the minimum set of seeds that spans the set of observed program behavior.

Previous work has formalized distillation as an instance of the
\emph{minimum set cover problem} (MSCP)~\cite{abdelnur:inria-00452015,
  andrey:inria-00546964, Rebert:2014}. MSCP is NP-complete (as also is
its weighted variant WMSCP)~\cite{Karp:1975:CCC:3160092.3160096}, so a
\emph{greedy} algorithm is usually applied to find an approximate
solution~\cite{10.2307/3689577}.

Thus, corpus distillation can be formalized as (W)MSCP, where the
``universe'' to be covered consists of code coverage information for
the set of seeds in the original collection corpus.  Code coverage is
conventionally used to characterize seeds in a fuzzing corpus due to
the strong positive correlation between code coverage and bugs found
while fuzzing~\cite{7081877, Gopinath:2014, Miller2008, Nagy2010}.
Finding the minimum set cover~\coverset is therefore equivalent to
finding the minimum set of seeds that still maintains the code
coverage observed in the collection corpus.  By definition,~\coverset
satisfies
\crefrange{corpus-prop:coverage}{corpus-prop:min-corpus-size} listed
in \cref{sec:Intro}.  Solving WMSCP, where weights correspond to the
seed size, also satisfies \cref{corpus-prop:min-seed-size}.


\section{Corpus Distillation Techniques}\label{sec:techniques}

\citet{abdelnur:inria-00452015} first introduced the idea of computing~\coverset
over code coverage as a seed selection strategy.  They used
a simple greedy algorithm to solve the unweighted MSCP.  Since then, a
number of corpus distillation techniques have been proposed.  The
remainder of this section presents these techniques, including our own
\tool approach to corpus distillation.

\subsection{\minset}
\citet{Rebert:2014} extended the work of
\citet{abdelnur:inria-00452015} by also computing~\coverset weighted
by execution time and file size.  They designed six corpus
distillation techniques and both simulated and empirically evaluated
these techniques over a number of fuzzing campaigns.
\citet{Rebert:2014} found that \uminset---an unweighted greedy-reduced
distillation---performed best in terms of distillation ability, and
that the \textsc{Peach Set} algorithm (based on the Peach fuzzer's
\texttt{peachminset} tool~\cite{Peach}) found the highest number of
bugs.  Curiously, \citet{Rebert:2014} also found
that \texttt{peachminset} does not in fact calculate the minimum cover
set, nor a proven competitive approximation thereof.  Our work extends
\citet{Rebert:2014} with a new theory, more extensive evaluation based on
modern \emph{coverage-guided greybox fuzzing}, and a more rigorous bug triage
process.

\subsection{\cmin}
Due to AFL's popularity, \cmin~\cite{AFL:2015} is perhaps the most widely-used
corpus distillation tool.  It implements a greedy distillation algorithm, but
has a unique approach to coverage.  In particular, \cmin reuses AFL's
own notion of edge coverage to categorize seeds at distillation time,
recording an approximation of edge \emph{frequency count}, not just
whether the edge has been taken.  When distilling, \cmin chooses the
smallest seed in the collection corpus that covers a given edge, and
then performs a greedy, weighted distillation.  We consider \cmin and
Rebert's \minset as representatives of the state-of-the-art in corpus
distillation tools, and include both in our evaluation.

\subsection{MoonShine}
MoonShine~\cite{Pailoor2018} is a corpus distillation tool for OS
fuzzers.  OS fuzzers typically test the system-call interface between
the OS kernel and user-space applications.  As such, the seeds that
are distilled by MoonShine are a list of system calls gathered from
program traces.  Our evaluation targets file-format fuzzing, which is a
fundamentally different problem to distilling system calls, and thus we do not
consider MoonShine in our evaluation.

\subsection{SmartSeed}
SmartSeed~\cite{Lv2018} takes a different approach from these others.
Rather than distilling a corpus of seeds, SmartSeed instead uses a
machine learning model to generate ``valuable'' seeds, where a seed is
considered valuable if it uncovers new code or produces a crash.
SLF~\cite{You:2019:SLF} takes this even further by producing valid
seeds from scratch by extracting information from the underlying
fuzzing infrastructure.

\subsection{\tool}
Our corpus distillation technique.  \tool represents the coverage data
for a corpus as a matrix: each row is a bit vector corresponding to
one seed, and each column to a possible \emph{edge} between basic
blocks in the target program (or library).  Such a matrix~$A$ has~$A_{ij} = 1$
if seed~$i$ causes the target to traverse edge~$j$, and is zero otherwise.

Following current state-of-the-art fuzzers (in particular, AFL), we
make the assumption that edge coverage is a good approximation of
target behavior, and thus: fuzzing over a distilled corpus will
discover as many bugs as fuzzing over the collection corpus.  Given
this, the objective is to find~$\mathit{minset}(A)$: the smallest
weighted set of seeds that covers all of the columns (edges) in~$A$ that have at
least one non-zero value.  (A column in~$A$
consisting of all zeroes represents an edge never taken by any of the
seeds.)

The unweighted version of the problem (MSCP as defined in
\cref{sec:FormalCorpusDistillation}) simply finds the smallest set of
rows (i.e., seeds) that spans all of the columns.  Much like \minset, \tool also
supports distillations weighted by \emph{file size} and \emph{execution time}.

To solve the (W)MSCP, the \tool algorithm applies dynamic programming
to take a large coverage matrix and recursively transform it through
row and column eliminations into successively smaller matrices while
accumulating a minimum cover set~\coverset.  These matrix operations include:
\begin{description}[nosep]
\item[Singularities:] Matrix rows/columns that sum to \emph{zero}. Row
  singularities represent seeds that do not cover any code when parsed
  by the target, while column singularities are an artifact of tracing
  tools that identify \emph{all} edges in the target.

  Singularities can be eliminated to produce a smaller matrix with the
  same~\coverset as~$A$.
\item[Exotic rows:] A row in~$A$ that is the only row covering a
  particular edge.  All seeds associated with exotic rows are by
  definition a part of the final solution and will be included in the
  distilled corpus.
\item[Dominant rows:] Row dominance captures the idea that some rows in~$A$ may
  be a subset of a single row.  The larger row
  \emph{dominates} the smaller \emph{submissive} row which is a subset
  of the \emph{dominator}.  In the MSCP, all submissive rows can be
  deleted from~$A$.  However, in the WMSCP, a submissive row can only
  be deleted if it has a larger weight than the dominator.
\item[Dominant columns:] Similar to row dominance, except the dominant
  column is deleted.  This is because any final solution by definition
  will contain seeds that cover the submissive columns and by
  implication will also cover the dominant column.
\item[Contained columns:] Eliminate columns that a chosen row (i.e.,
  seed) covers.  The columns can be safely deleted because they will
  be covered by the seed associated with the chosen row.
\item[Heuristic rows:] The previously-described operations have been
  \emph{optimal} in the sense that they guarantee an optimal solution
  for a smaller transformed matrix can be used to construct an optimal
  solution for the larger matrix. However, in the case where an
  optimal operation cannot be made, \tool must make a \emph{heuristic
  choice} to select a row to add to~\coverset.  In the MSCP, a good
  heuristic is to select the row with the largest row sum. In the
  WMSCP, we choose the row with the largest \emph{weighted} row sum.
\end{description}

\tool is open-source and freely available at \bitlytools.  In addition to \tool,
we also provide \emph{\bvtool}, a tool that generates bit vector traces for all
seeds in the collection corpus. \bvtool converts the output of \showmap (a tool
included with AFL to display the coverage trace of a particular input) to the
bit vector representation used by \tool.  The output of \showmap is also used by
\cmin.

\subsection{Motivating Weighted Corpus Distillations}
Industrial-scale fuzzing involves a large number of worker processes campaigning
on a given target.  For example, Google reports that ClusterFuzz runs \num{5000}
fuzzers on over~\num{25000} cores, churning \emph{4~trillion} test-cases a
week~\cite{Google:2019:ClusterFuzzBlackHat, Google:2019:ClusterFuzzBlog,
OSSFuzz:2016}.  This places a large I/O burden on the fuzzing infrastructure, as
test-cases must be fetched/loaded from the global corpus, saved to the (local)
fuzzing queue (when new code is discovered), and synchronized with the global
corpus (so that new coverage can be shared with the other worker
processes).\footnote{Not all fuzzers synchronize with an explicit global corpus.
Instead, they may synchronize with the other worker processes' queues directly.}

Previous work has demonstrated the impact that file system contention
has on industrial-scale fuzzing~\cite{Xu:2017}: despite fuzzing being
embarrassingly-parallel, the number of test-case executions saturates
at~15 cores, degrades at~30 cores, and collapses at~120 cores.  This
collapse is due to overhead from opening/closing test-cases ($2\times$
slowdown) and queue syncing between workers (a further~$2\times$
overhead)~\cite{Xu:2017}.  In our experiments, we found that syncing
accounted for~\SI{63.78}{\percent} of the operations that wrote to the
fuzzing queue (and hence to the file system).  As noted by
\citet{Xu:2017}, this time spent syncing (hence re-executing
test-cases from previous worker processes) is time diverted from
mutating inputs and expanding coverage.  Therefore, a weighted corpus
distillation, minimizing the total collective byte size of the fuzzing
corpus, alleviates the I/O demand on the storage system.  Given this,
practical fuzzing would seem to benefit most from using file-size
weighted distillations compared to unweighted.


\section{Evaluation Methodology}\label{sec:Evaluation}

We evaluate \numdistills different corpus design approaches which we
shortly describe in detail.  We show that corpus distillation has
significant impact for long fuzzing campaigns.  Notably, we find that
neither of the two state-of-the-art approaches---\tool and
\minset---are able to find all of the bugs that we discovered in our
target set.  However, there are a range of other conclusions we make
based on the large number of experiments we have conducted.

\subsection{Experimental Setup}

Our experiments were conducted on a pair of identically configured
Dell Poweredge servers with~48-core Intel(R) Xeon(R) Gold~5118~2.30GHz
CPUs,~512GB of RAM, Hyper-Threading enabled (providing a total of~96
logical CPUs), and running Ubuntu~18.04.

\subsection{Target Selection}\label{sec:target-selection}

We use the Google Fuzzer Test Suite (FTS)~\cite{Google:FTS} and
\numrealprogs popular open-source programs (spanning \numfiletypes
different file formats) to test different corpus design approaches.
These targets are detailed in \cref{tab:Targets}.  We exclude some FTS
targets, because:
\begin{myinlinelist}
\item they contain only memory leaks (e.g., proj4-2017-08-14), which
  are not detected by AFL by default; or
\item we were unable to find a suitably-large collection corpus for a
  particular file type (e.g., ICC files for lcms-2017-03-21).
\end{myinlinelist}
This left us with \numftsprogs of the original~24 targets.  We elected
to use the FTS over the CGC~\cite{DARPA:2018} or
LAVA-M~\cite{LAVA:2016} benchmarks because CGC and LAVA-M:
\begin{myinlinelist}
\item do not resemble ``real world'' programs/bugs; and
\item mostly accept text as input, rather than a range of diverse
  binary formats.
\end{myinlinelist}
Furthermore, Google's FuzzBench~\cite{Google:FuzzBench} was not used
because it was not available at the time of writing.  However,
thirteen of the 24 FuzzBench targets also exist in the FTS.  Of these
thirteen targets, we include six in our evaluation; the remaining
seven are excluded for the reasons given previously.  Of the eleven
FuzzBench targets not included in the FTS, four are also unsuitable
(for the same reasons).

The~\numrealprogs real-world targets (\cref{tab:AdditionalTargets})
were selected to be representative of popular programs that are
commonly fuzzed and that operate on a diverse range of file formats
(e.g., images, audio, and documents).  The driver program used for
each target library is shown in
parentheses.\footnote{The driver \texttt{char2svg} was adapted from
  \url{https://www.freetype.org/freetype2/docs/tutorial/example5.cpp}.}

\begin{table}
    \centering
    \scriptsize
    \caption{Fuzzing targets.}
    \label{tab:Targets}
    \vspace{-1em}
    \begin{subtable}[t]{0.45\textwidth}
      \centering
      \caption{Google Fuzzer Test Suite targets.}
      \label{tab:GoogleFTSTargets}

      \rowcolors{2}{white}{rowgray}
      \begin{tabular}{ll|ll}
      \toprule
      Program & File type & Program & File type \\
      \midrule
      freetype2-2017 & TTF &
      guetzli-2017-3-30 & JPEG \\
      json-2017-02-12 & JSON &
      libarchive-2017-01-04 & GZIP \\
      libjpeg-turbo-07-2017 & JPEG &
      libpng-1.2.56 & PNG \\
      libxml2-v2.9.2 & XML &
      pcre2-10.00 & Regex \\
      re2-2014-12-09 & Regex &
      vorbis-2017-12-11 & OGG \\
      \bottomrule
      \end{tabular}
    \end{subtable}
    \\[1em]
    \begin{subtable}[t]{0.45\textwidth}
      \centering
      \caption{Real-world targets.}
      \label{tab:AdditionalTargets}
      \rowcolors{2}{white}{rowgray}
      \begin{tabular}{lrl}
      \toprule
      Program (driver) & Version & File type \\
      \midrule
      Poppler (\texttt{pdftotext})    & 0.64.0  & PDF \\
      SoX (\texttt{sox})              & 14.4.2  & MP3 \\
      SoX (\texttt{sox})              & 14.4.2  & WAV \\
      librsvg (\texttt{rsvg-convert}) & 2.40.20 & SVG \\
      libtiff (\texttt{tiff2pdf})     & 4.0.9   & TIFF \\
      FreeType (\texttt{char2svg})    & 2.5.3   & TTF \\
      libxml2 (\texttt{xmllint})      & 2.9.0   & XML \\
      \bottomrule
      \end{tabular}
    \end{subtable}
\end{table}

\subsection{Sample Collection}

For each file type in \cref{sec:target-selection}, we built a Web
crawler using Scrapy\footnote{\url{https://scrapy.org/}} to crawl the
Internet for~72~hours to create the collection corpus.  For image
files, crawling started with Google search results and the Wikimedia
Commons repository.  For media and document files (e.g., PDF),
crawling started from the Internet Archive and Creative Commons
collections.  The regular expressions used in pcre2 and re2 were
obtained from regexlib,\footnote{\url{http://regexlib.com/}} while OGG
files were sourced from old video games%
\footnote{\url{www.kenney.nl},
  \url{https://www.themotionmonkey.co.uk/},
  \url{https://opengameart.org/}}%
(in addition to the Internet Archive).  Finally, we found TIFF files
to be relatively rare, so~\SI{40}{\percent} of the TIFF seeds were
generated by converting other image types such as JPEG and BMP using
ImageMagick.

We preprocessed each collection corpus to remove duplicates identified
by MD5 checksum, and files larger than~\SI{300}{\kibi\byte}.  The
cutoff file size~\SI{300}{\kibi\byte} is our best effort to conform to
the AFL authors' suggestions regarding seed size, while still having
enough eligible seeds in the preprocessed corpora.  We split audio
files larger than~\SI{1}{\mebi\byte} into smaller files using FFmpeg.
In total, we collected \numcollectedseeds seeds across \numfiletypes
different file formats.  After preprocessing our collection corpus we
were left with a total of \numcandidateseeds seeds.

\subsection{Fuzzer Setup}

We ran one fuzzing \emph{campaign} per target/file-type per
distillation technique.  Each fuzzing campaign consists of thirty
independent 18 hour \emph{trials}.  We emphasize the large number of
repeated trials here because we found (consistent with
\citet{Hicks:2018}) that individual fuzzing trials vary wildly in
performance. Therefore, reaching statistically meaningful conclusions
requires many trials.  The length of each trial and the number of
repeated trials satisfy the recommendations presented by
\citet{Hicks:2018}.

We configure AFL~(v2.52b) for single-system parallel
execution\footnote{\url{https://github.com/mirrorer/afl/blob/master/docs/parallel_fuzzing.txt}}
with one master and several secondary nodes.  When evaluating the FTS,
we used a single secondary node (allowing one node to focus on
deterministic checks, while the other node proceeds straight to havoc
mode).  When fuzzing the real-world targets, we scaled up to seven
secondary nodes.  However, in practice we found this to be futile, as
the seven secondary nodes tended to behave similarly in the scheduling
of inputs that they fuzzed.

We compile each target using AFL's LLVM (v7) instrumentation for~32-bit
\texttt{x86} with Address Sanitizer (ASan)~\cite{ASAN} enabled.
We chose LLVM instrumentation over AFL's assembler-based
instrumentation because it offers the best level of interoperability
with ASan.

We tuned AFL's available virtual memory parameter for each target to
enable effective fuzzing.\footnote{Per-target settings are available
at \bitlydata.}  When fuzzing the FTS we configured the target
process to respawn after \emph{every} iteration.  This was due to
stability issues that we encountered when fuzzing in single-system
parallel execution mode.  All other AFL parameters were left at their
default values.

\subsection{Experiment}\label{sec:experiment}

We evaluate \numdistills distillation techniques (see below) against
the targets described in \cref{sec:target-selection}.  For each
distillation technique we perform thirty distinct trials of~\SI{18}{\hour}
of fuzzing per trial using the same distilled corpus.
In total, this amounts to~\num{3180} individual trials and over~\numcpuyears of
fuzzing, providing ample empirical support for the
simulation-based analyses undertaken by \citet{Rebert:2014}.

We compared the following distillation techniques:
\begin{description}[nosep]
\item[Full] The collection corpus without distillation, preprocessed to remove
  duplicates and filtering for size, as previously discussed.

\item[CMIN] AFL's \cmin tool for corpus distillation. 

\item[MS-U] The \uminset tool. 
  We present \uminset (as opposed to \textsc{Time} or \textsc{Size} \minset) because it finds the most bugs of the various \minset configurations~\cite{Rebert:2014}.

\item[ML-S] The \tool algorithm \emph{weighted by file size}.

\item[Empty] We also evaluate a corpus
  for each target comprising just an ``empty'' seed, following
  \citet{Hicks:2018} who reported that \emph{``despite its use contravening
  conventional wisdom,''} the empty seed outperformed (in terms of bug yield) a
  set of valid non-empty seeds for some targets~\cite{Hicks:2018}. Our
  ``empty'' seed is not merely a zero-length input, but rather a small file
  handcrafted to contain the bytes necessary to satisfy file header checks.
  More details on these ``empty'' seeds can be found in \cref{app:EmptySeeds}.
\end{description}
We also explored a random sampling of the collection (Full) corpus
(following \citet{Rebert:2014}), in addition to unweighted and
execution-time weighted variants of \tool.  However, these
distillation techniques performed poorly (compared to the \numdistills
techniques listed above) and so we omit these results.  All raw data
(including for the omitted results) is available at \bitlydata.

We compare the performance of each distillation technique across four
measures:
\begin{description}[nosep]
\item[Code coverage] Coverage is often used to measure fuzzing
  effectiveness, as ``\textit{covering more code intuitively
    correlates with finding more bugs}''~\cite{Hicks:2018}.  We use
  coverage as reported by AFL.
\item[Bug count] While code coverage is a common metric, its
  correlation with bug-finding effectiveness may be
  weak~\cite{Inozemtseva:2014:CoverageCorrelation}.  Therefore, a
  direct bug count is preferable for comparing fuzzer
  effectiveness~\cite{Hicks:2018}.  To this end, we perform manual
  triage for \emph{all} fuzzer-produced crashes, isolating the bugs
  that led to those crashes.  This is in contrast to much of the
  existing literature~\cite{Rebert:2014,Hicks:2018,Wang:2019:AFLSensitive,Li:2019:Cerebro},
  which uses stack-hash deduplication to determine unique bugs from crashes,
  a technique known to both over \emph{and} under count
  bugs~\cite{Hicks:2018}.
\item[Bug-finding reliability] As discussed earlier, fuzzing is a
  highly stochastic process, and individual trials vary wildly in
  bug-finding performance.  As such, we also measure how reliable a
  corpus is at uncovering a particular bug.  We do this by counting
  the number of times an individual trial found a given bug.
\item[Time-to-bug] The faster a bug is found and reported, the quicker
  it can be fixed.  To this end, we also report the time until first
  discovery of a given bug.  This is calculated as the arithmetic mean
  of the time taken to find the bug for those trials that successfully
  found it (we omit those trials that fail to find the bug).
\end{description}


\section{Results}

\begin{table}
  \centering
  \fontsize{6pt}{7pt}\selectfont
  \caption{Comparison of corpora for both benchmark suites.
Each corpus is summarized by its \emph{number of files} (``\#'') and  \emph{total size} (``S'')---summing the sizes of all included files~(MB).
Cell colour denotes the best performing technique: blue for ``\#'' and green for ``S'' (ties are not included).
}
  \label{tab:CorpusSizes}

  \begin{tabular}{p{7mm}|%
                  >{\raggedleft\arraybackslash}p{7mm}>{\raggedleft\arraybackslash}p{7mm}|%
                  >{\raggedleft\arraybackslash}p{5mm}>{\raggedleft\arraybackslash}p{5mm}|%
                  >{\raggedleft\arraybackslash}p{4mm}>{\raggedleft\arraybackslash}p{4.5mm}|%
                  >{\raggedleft\arraybackslash}p{4mm}>{\raggedleft\arraybackslash}p{4.5mm}}
    \toprule
                 & \multicolumn{2}{c|}{Full} & \multicolumn{2}{c|}{CMIN} & \multicolumn{2}{c|}{MS-U} & \multicolumn{2}{c}{ML-S} \\
    Target & \#  & S  & \#  & S  & \#  & S  & \#  & S     \\
    \midrule
    \multicolumn{9}{l}{\textbf{Google FTS}} \\
    \rowcolor{rowgray}
    freetype2     & \num{466}         & \num{35.50}      & \num{246}         & \num{20.91}      & \num{43}          & \num{5.40}       & \bcell \num{42}          & \gcell \num{5.23}       \\
    guetzli       & \num{120000}      & \num{222.85}     & \num{463}         & \num{0.59}       & \num{17}          & \num{0.04}       & \bcell \num{16}          & \gcell \num{0.02}       \\
    \rowcolor{rowgray}
    json          & \num{19978}       & \num{76.45}      & \num{149}         & \num{2.56}       & \bcell \num{17}          & \num{0.95}       & \num{25}          & \gcell \num{0.52}       \\
    libarchive    & \num{108558}      & \num{850.64}     & \num{180}         & \num{2.79}       & \bcell \num{41}          & \num{3.18}       & \num{46}          & \gcell \num{0.73}       \\
    \rowcolor{rowgray}
    libjpeg-turbo & \num{120000}      & \num{222.85}     & \num{93}          & \num{0.10}       & \bcell \num{3}           & \num{0.01}       & \num{5}           & \num{0.01}                           \\
    libpng        & \num{66512}       & \num{7773.60}    & \num{107}         & \num{4.05}       & \bcell \num{22}          & \num{1.91}       & \num{25}          & \gcell \num{1.14}       \\
    \rowcolor{rowgray}
    libxml2       & \num{79032}       & \num{205.64}     & \num{440}         & \num{7.70}       & \bcell \num{97}          & \num{2.23}       & \num{113}         & \gcell \num{0.65}       \\
    pcre2         & \num{4520}        & \num{0.45}       & \num{691}         & \num{0.13}       & \bcell \num{183}         & \num{0.04}       & \num{188}         & \gcell \num{0.03}       \\
    \rowcolor{rowgray}
    re2           & \num{4520}        & \num{0.45}       & \num{155}         & \num{0.01}       & \num{56}          & \num{0.01}       & \bcell \num{54}          & \num{0.01}                           \\
    vorbis        & \num{99450}       & \num{8902.70}    & \num{237}         & \num{12.06}      & \bcell \num{8}           & \num{0.30}       & \num{9}           & \gcell \num{0.10}       \\
    \midrule
    \multicolumn{9}{l}{\textbf{Real-world Targets}} \\
    \rowcolor{rowgray}
    Poppler       & \num{99984}       & \num{6086.70}    & \num{1318}        & \num{121.90}     & \bcell \num{189}         & \num{22.70}      & \num{209}         & \gcell \num{17.32}      \\
    SoX (MP3)     & \num{99691}       & \num{4094.40}    & \num{147}         & \num{3.75}       & \bcell \num{9}           & \num{0.17}       & \num{11}          & \gcell \num{0.09}       \\
    \rowcolor{rowgray}
    SoX (WAV)     & \num{74000}       & \num{2490.60}    & \num{68}          & \num{1.65}       & \bcell \num{10}          & \num{0.39}       & \num{11}          & \gcell \num{0.26}       \\
    librsvg       & \num{71763}       & \num{744.59}     & \num{881}         & \num{17.05}      & \bcell \num{173}         & \num{4.34}       & \num{183}         & \gcell \num{2.58}       \\
    \rowcolor{rowgray}
    libtiff       & \num{99955}       & \num{466.52}     & \num{67}          & \num{0.27}       & \bcell \num{23}          & \num{0.10}       & \bcell \num{23}          & \gcell \num{0.09}       \\
    FreeType      & \num{466}         & \num{35.50}      & \num{73}          & \num{8.68}       & \bcell \num{23}          & \num{3.04}       & \bcell \num{23}          & \gcell \num{2.92}       \\
    \rowcolor{rowgray}
    libxml2       & \num{79032}       & \num{205.64}     & \num{505}         & \num{9.04}       & \bcell \num{103}         & \num{1.67}       & \num{120}         & \gcell \gcell \num{0.96}       \\
    \bottomrule
    \end{tabular}
\end{table}

\cref{tab:CorpusSizes,tab:fts-results,tab:triage-results} and
\cref{fig:fuzzing-bugs} summarize our experimental results.
\cref{tab:CorpusSizes} displays the distillation results for the
corpora, with the best-performing corpora highlighted.
It can be seen that corpora produced by both ML-S and MS-U
are always smaller (by~\SIrange{80}{82}{\percent} in terms of the
number of files and~\SIrange{67}{78}{\percent} in terms of total size)
than that produced by CMIN.
While it does not make sense to compare ML-S with MS-U since they optimize
for different objectives (the former is weighted, while the latter is
unweighted), we still observe that ML-S outperforms or is equal to MS-U five out
of~17 times on MS-U's own optimization objective.  The reverse is never true.

What ultimately matters is if the distilled corpora lead to better
fuzzing outcomes.  To this end, we now discuss the bug-finding ability
of the different corpus distillation techniques across our two
benchmark suites (the Google FTS and a set of real-world targets).  We
analyze these results with respect to the performance measures
outlined in \cref{sec:experiment}.

\subsection{Google Fuzzer Test Suite}\label{sec:google-fts-evaluation}

\begin{table*}[h]
  \centering
  \caption{Google FTS results.  These results include three metrics:
    the number of times a bug was found (``\#''); the mean (with
    standard deviation) of time-to-bug in hours (``T''); and mean code
    coverage (``C'') for each corpus.  Bug IDs are derived from the
    order in which the bugs are presented in the target's README (from
    the FTS repo).  Bugs marked with \textdagger\ denote benchmarks
    that attempt to verify that the fuzzer can reach a known location.
    The best performing corpus for each target in terms of number of
    times the bug was found, mean time-to-bug, and mean code coverage
    is highlighted in yellow, green, and blue respectively (ties are
    not included).  }
  \label{tab:fts-results}

  \begin{adjustbox}{width=0.95\textwidth}
    \begin{tabular}[t]{p{3.5em}|l|rrr|rrr|rrr|rrr|rrr|rrr}
    \toprule
    & & \multicolumn{3}{c|}{FTS} & \multicolumn{3}{c|}{CMIN} & \multicolumn{3}{c|}{ML-S} & \multicolumn{3}{c|}{MS-U} & \multicolumn{3}{c|}{Full} & \multicolumn{3}{c}{Empty} \\
        \multirow{-2}{*}{Target} & \multirow{-2}{3mm}{Bug ID} & \# & T (h) & C (\%) & \# & T (h) & C (\%) & \# & T (h) & C (\%) & \# & T (h) & C (\%) & \# & T (h) & C (\%) \\
    \midrule
    \rowcolor{rowgray}
    freetype2 & A\textsuperscript{\textdagger} & 27 & $6.14 \pm 5.07$ & $14.46$ & $30$ & $0 \pm 0$ & \bcell\num{17.93} & $30$ & $0 \pm 0$ & \num{17.20} & $30$ & $0 \pm 0$ & \num{17.29} & $30$ & $0 \pm 0$ & \num{17.52} & 0 & N/A & \num{7.78} \\
        guetzli & A & \ycell 23 & $7.89 \pm 4.82$ & \bcell $7.27$ & $2$ & $13.68 \pm 4.01$ & \num{7.13} & $11$ & \gcell$7.98 \pm 5.77$ & \num{7.14} & $10$ & $11.57 \pm 3.91$ & \num{7.06} & $0$ & N/A & \num{6.19} & 0 & N/A & \num{2.40} \\
    \rowcolor{rowgray}
        json & A & \ycell 30 & \gcell $0.06 \pm 0.09$ & $2.13$ & $2$ & $1.96 \pm 0.70$ & \num{2.15} & $1$ & $0.82$ & \num{2.15} & $1$ & $7.30$ & \num{2.15} & 0 & N/A & \num{2.12} & 0 & N/A & \num{2.12} \\
        libarchive & A & 0 & N/A & $4.75$ & $30$ & $9.38 \pm 4.09$ & \bcell\num{4.95} & $30$ & $13.32 \pm 0.78$ & \num{4.89} & $30$ & \gcell$4.43 \pm 0.59$ & \num{4.94} & 0 & N/A & \num{4.49} & 0 & N/A & \num{4.86} \\
    \rowcolor{rowgray}
        libjpeg-turbo & A\textsuperscript{\textdagger} & 30 & $3.09 \pm 2.71$ & $3.86$ & $30$ & $3.79 \pm 3.43$ & \num{4.09} & $30$ & $2.93 \pm 2.73$ & \num{4.09} & $30$ & $3.34 \pm 2.84$ & \num{4.09} & $3$ & $8.27 \pm 4.59$ & \num{3.09} & $30$ & \gcell$1.90 \pm 1.31$ & \num{3.66} \\
     & A\textsuperscript{\textdagger} & 30 & $0.08 \pm 0.21$ & & $30$ & $0 \pm 0$ & \bcell & $30$ & $0 \pm 0$ & & $30$ & $0 \pm 0$ & & $30$ & $0 \pm 0$ & & $30$ & $0 \pm 0$ & \\
     & B\textsuperscript{\textdagger} & 30 & $0 \pm 0$ & & $30$ & $0 \pm 0$ & \bcell & $30$ & $0 \pm 0$ & & $30$ & $0 \pm 0$ & & $30$ & $0 \pm 0$ & & $0$ & N/A & \\
    \multirow{-3}{*}{libpng} & C\textsuperscript{\textdagger} & 30 & $0.01 \pm 0.00$ & \multirow{-3}{*}{$1.43$} & $30$ & $0 \pm 0$ & \multirow{-3}{*}{\bcell\num{2.03}} & $30$ & $0 \pm 0$ & \multirow{-3}{*}{\num{2.02}} & $30$ & $0 \pm 0$ & \multirow{-3}{*}{\num{2.01}} & $30$ & $0 \pm 0$ & \multirow{-3}{*}{\num{1.86}} & $30$ & $0 \pm 0$ & \multirow{-3}{*}{\num{1.21}} \\
    \rowcolor{rowgray}
        & A & \textminus & \textminus & & $30$ & $0.77 \pm 0.35$ & & $30$ & $0.65 \pm 0.20$ & \bcell & $30$ & \gcell$0.53 \pm 0.16$ & & $30$ & $3.00 \pm 0.82$ & & $0$ & N/A & \\
    \rowcolor{rowgray}
        & B & \textminus & \textminus & & $23$ & $8.30 \pm 4.05$ & & $24$ & $8.59 \pm 4.22$ & \bcell & $16$ & $9.27 \pm 3.90$ & & $12$ & $12.65 \pm 3.19 $ & & \ycell$29$ & \gcell$4.89 \pm 2.64$ & \\
    \rowcolor{rowgray}
        \multirow{-3}{*}{libxml2} & C & \textminus & \textminus & \multirow{-3}{*}{\textminus} & $1$ & \gcell$3.77$ & \multirow{-3}{*}{\num{14.49}} & \ycell$4$ & $10.55 \pm 5.31$ & \multirow{-3}{*}{\bcell\num{14.58}} & 0 & N/A & \multirow{-3}{*}{\num{14.42}} & $0$ & N/A & \multirow{-3}{*}{\num{13.67}} & $0$ & N/A & \multirow{-3}{*}{\num{5.96}} \\
        & A & \textminus & \textminus & & $30$ & $2.07 \pm 0.69$ & \bcell & $30$ & $2.46 \pm 1.10$ & & $30$ & $2.15 \pm 0.69$ & & $30$ & $2.54 \pm 1.12$ & & $30$ & \gcell$1.88 \pm 0.73$ & \\
        \multirow{-2}{*}{pcre2}  & B & \textminus & \textminus & \multirow{-2}{*}{\textminus} & $30$ & $2.29 \pm 1.89$ & \multirow{-2}{*}{\bcell\num{10.21}} & $30$ & $2.40 \pm 2.24$ & \multirow{-2}{*}{\num{10.18}} & $30$ & \gcell$2.01 \pm 1.97$ & \multirow{-2}{*}{\num{10.19}} & $30$ & $2.01 \pm 2.08$ & \multirow{-2}{*}{\num{10.17}} & $30$ & $3.25 \pm 2.03$ & \multirow{-2}{*}{\num{9.93}} \\
    \rowcolor{rowgray}
        & A\textsuperscript{\textdagger} & \textminus & \textminus & & $30$ & \gcell$0.52 \pm 0.36$ & & $30$ & $0.68 \pm 0.54$ & & $30$ & $0.74 \pm 1.51$ & & $30$ & $0.82 \pm 0.47$ & & $30$ & $1.73 \pm 0.91$ & \bcell \\
    \rowcolor{rowgray}
        \multirow{-2}{*}{re2} & B & \textminus & \textminus & \multirow{-2}{*}{\textminus} & $16$ & $6.47 \pm 5.59$ & \multirow{-2}{*}{\num{6.76}} & $10$ & $6.96 \pm 5.14$ & \multirow{-2}{*}{\num{6.76}} & \ycell$18$ & $7.21 \pm 4.20$ & \multirow{-2}{*}{\num{6.76}} & $4$ & \gcell$4.43 \pm 4.23$ & \multirow{-2}{*}{\num{6.74}} & $2$ & $12.97 \pm 4.58$ & \multirow{-2}{*}{\num{6.80}} \bcell \\
    \bottomrule
  \end{tabular}
  \end{adjustbox}
\end{table*}

\Cref{tab:fts-results} summarizes the bugs found in the Fuzzer Test Suite (FTS).
We conduct an additional campaign, FTS, with the seeds provided by the FTS developers (for targets where seeds are provided; the libxml2, pcre2, and re2 targets did not have seeds).
The vorbis target is omitted because none of its three bugs were found with any of the corpora.

Notably, four of the six coverage benchmarks are reached instantaneously (i.e., seeds in the fuzzing corpora reach the particular line of code without requiring any fuzzing) by all corpora \emph{except} FTS and EMPTY.
Naturally, EMPTY takes some time to reach the target locations, as AFL must construct valid inputs from ``nothing''.
Nevertheless, EMPTY completes four of the six coverage benchmarks within two hours (on average).\footnote{Two of the libpng benchmarks are reached instantaneously, even with EMPTY.
This is because our empty PNG (see \cref{app:EmptySeeds}) contains the required PNG elements---an IHDR header and compressed IDAT chunk---to reach the code location.}
The freetype2 and a libpng locations are never reached by EMPTY, because:
\begin{description}
\item[freetype2] requires a composite glyph, which EMPTY never produces; and

\item[libpng] requires a specific chunk type (sRGB), which is difficult to
synthesize without any knowledge of the PNG file format.
\end{description}

The two benchmarks that are not reached instantaneously---re2
and libjpeg-turbo---are reliably reached by all corpora \emph{except}
FULL within the first four hours (on average) of each trial.  The FULL
corpus is highly unreliable on libjpeg-turbo: it only reaches the
target location in~\SI{10}{\percent} of trials, and when it does, it
takes double the time of the other corpora.

These results demonstrate both the benefit of maximizing code coverage upfront
and minimizing duplicate behavior (per
\cref{corpus-prop:coverage,corpus-prop:redundant-seeds} respectively): the
fuzzer does not have to rely solely on random mutation to uncover new program
behaviors, and redundant seeds are wasteful and clog the fuzzing queue.  The
benefit of maximizing code coverage upfront is reinforced by EMPTY's coverage
statistics: it achieves the lowest mean code coverage in six of the ten FTS
targets.

Of the eight benchmarks that EMPTY was able to complete, it was the
(equal) fastest to do so for five of these.  However, it suffers from
the highest ``false negative'' rate: i.e., it is the most likely
corpus to miss a bug when one exists (as evident from the number of
``N/A'' entries in \cref{tab:fts-results}).  We hypothesize that this
speed is due to the reduced search space when mutating the empty seed,
but that the mutation engine is less likely to ``get lucky'' in
generating a bug-inducing input when starting from nothing.
Conversely, FULL was the slowest on all but one target (re2).

The seeds provided in the FTS generally perform well. In particular, the
provided json seed is faster and more reliable than the three distilled corpora.
This is unsurprising, as the json bug is known to be ``\textit{found in about
five minutes using the provided seed}''~\cite{Google:FTS}. However, the
performance of the other five targets was worse when using the provided FTS
seeds. In particular, the libarchive bug was never found by the FTS seed; in
comparison, this bug was found reliably by \emph{all} distilled corpora (CMIN,
MLS-S, and MS-U).

Finally, guetzli's results are worth further discussion.  With the
exception of EMPTY, guetzli achieves a relatively low number of
executions per second ($\sim 6$ executions per second).  This low
iteration rate has the largest impact on the FULL corpus: AFL is not
able to complete an initial pass over the~\num{120000} seeds in this
corpus (in an~\SI{18}{\hour} trial), let alone perform any mutations
and discover the bug in this target.  This highlights the need for
distillation when starting with a large collection corpus (i.e., the
importance of
\cref{corpus-prop:min-corpus-size,corpus-prop:min-seed-size}): all
three distilled corpora (CMIN, ML-S, and MS-U) and FTS were able to find the
guetzli bug within similar time frames.

\subsection{Real-World Targets}\label{sec:additional-targets-evaluation}

\begin{table*}
  \centering
  \scriptsize
  \caption{Real-world target results.  These results include two
    metrics: the number of times a bug was found (``\#''); and the
    mean (with standard deviation) of time-to-bug (``T'') for each
    corpus.  The best performing corpus for each target in terms of
    mean number of bugs found and relative bug-finding speed is
    highlighted in yellow and green respectively (ties are not
    included).  }
  \label{tab:triage-results}

  \begin{subtable}[t]{0.49\textwidth}
  \centering
    \begin{adjustbox}{width=0.95\textwidth}
    \begin{tabular}[t]{p{0.5em}p{3em} rrrrr l}
    \toprule
    \rotatebox{90}{Bug ID} & \rotatebox{90}{Metric} & \rotatebox{90}{CMIN} & \rotatebox{90}{ML-S} & \rotatebox{90}{MS-U} & \rotatebox{90}{Full} & \rotatebox{90}{Empty} & CVE \\
    \midrule
    \multicolumn{8}{l}{\textbf{libxml2}} \\
      \rowcolor{rowgray} & \# & $30$ & $30$ & $30$ & $30$ & $2$ & \\
      \rowcolor{rowgray} A & & $0.51$ & \gcell$0.34$ & $0.36$ & $2.84$ & $5.19$ & 2015-8317 \\
        \rowcolor{rowgray} & \multirow{-2}{*}{T (h)} & $\pm0.14$ & \gcell$\pm0.20$ & $\pm0.10$ & $\pm0.71$ & $\pm5.23$ & \\
      \rowcolor{rowgray} \bugdesc{Heap buffer overread} \\
        & \# & $29$ & $27$ & $29$ & $28$ & \ycell$30$ &  \\
        B & & $5.89$ & $7.45$ & $6.91$ & $10.84$ & \gcell$1.49$ & 2015-7497 \\
        & \multirow{-2}{*}{T (h)} & $\pm 2.66$ & $\pm 4.42$ & $\pm 4.08$ & $\pm 2.15$ & \gcell$\pm 0.97$ & \\
      \bugdesc{Negative index into array} \\
      \rowcolor{rowgray} & \# & $30$ & $30$ & $30$ & $25$ & $0$ & \\
      \rowcolor{rowgray} C & & \gcell$1.82$ & $2.57$ & $2.72$ & $6.67$ & N/A & 2015-5312 \\
        \rowcolor{rowgray} & \multirow{-2}{*}{T (h)} & \gcell$\pm 0.64$ & $\pm 1.53$ & $\pm 1.96$ & $\pm 2.83$ & & \\
      \rowcolor{rowgray} \bugdesc{Denial of service} \\
      & \# & \ycell$4$ & $2$ & $2$ & $0$ & $0$ & \\
        D & & $10.35$ & \gcell$6.50$ & $15.01$ & N/A & N/A & 2016-1835 \\
        & \multirow{-2}{*}{T (h)} & $\pm 6.56$ & \gcell$\pm 0.32$ & $\pm 0.48$ & & & \\
      \bugdesc{Use-after-free} \\
      \rowcolor{rowgray} & \# & $1$ & \ycell$2$ & $0$ & $1$ & $0$ & \\
      \rowcolor{rowgray} E & & $10.38$ & $10.14$ & N/A & \gcell$4.76$ & N/A & 2016-1836 \\
        \rowcolor{rowgray} & \multirow{-2}{*}{T (h)} & & $\pm 2.19$ & & & & \\
      \rowcolor{rowgray} \bugdesc{Use-after-free} \\
      & \# & $10$ & \ycell$12$ & $3$ & $2$ & $0$ & \\
      F & & $10.05$ & \gcell$9.32$ & $14.67$ & $12.23$ & N/A & 2016-1762 \\
        & \multirow{-2}{*}{T (h)} & $\pm 4.97$ & \gcell$\pm 5.52$ & $\pm 3.76$ & $\pm 1.76$ & & \\
      \bugdesc{Continuation after error} \\
      \rowcolor{rowgray} & \# & \ycell$1$ & $0$ & $0$ & $0$ & $0$ & \\
      \rowcolor{rowgray} \multirow{-2}{*}{G} & & \gcell$10.13$ & N/A & N/A & N/A & N/A & \multirow{-2}{*}{2016-3627} \\
      \rowcolor{rowgray} \bugdesc{Infinite recursion} \\
      & \# & $0$ & \ycell$1$ & $0$ & $0$ & $0$ & \\
      \multirow{-2}{*}{H} & & N/A & \gcell$4.76$ & N/A & N/A & N/A & \multirow{-2}{*}{2015-7942} \\
      \bugdesc{Input buffer overread} \\
      \rowcolor{rowgray} & \# & $0$ & $0$ & $1$ & $0$ & $1$ & \\
      \rowcolor{rowgray} \multirow{-2}{*}{I} & & N/A & N/A & $3.29$ & N/A & \gcell$3.14$ & \multirow{-2}{*}{2015-7499} \\
      \rowcolor{rowgray} \bugdesc{Heap buffer overflow} \\
      & \# & $8$ & $5$ & $5$ & $9$ & \ycell$30$ & \\
      J & & $5.42$ & $11.57$ & $3.18$ & $7.47$ & \gcell$1.74$ & 2015-7498 \\
        & \multirow{-2}{*}{T (h)} & $\pm 4.24$ & $\pm 7.73$ & $\pm 2.73$ & $\pm 3.72$ & \gcell$\pm 0.98$ & \\
      \bugdesc{Heap buffer overflow} \\
    \midrule
    \multicolumn{8}{l}{\textbf{libtiff}} \\
      \rowcolor{rowgray} & \# & $11$ & $21$ & $12$ & $2$ & \ycell$26$ & \\
      \rowcolor{rowgray} A & & $7.69$ & $8.23$ & $8.27$ & $9.71$ & \gcell$4.18$ & 2019-14973 \\
        \rowcolor{rowgray} & \multirow{-2}{*}{T (h)} & $\pm 6.02$ & $\pm 6.15$ & $\pm 6.00$ & $\pm 4.96$ & \gcell$\pm 4.10$ & \\
      \rowcolor{rowgray} \bugdesc{Elision of integer overflow check by compiler} \\
      & \# & $3$  & $4$  & \ycell$5$ & $0$ & $0$ & \\
        B & & $16.04$ & \gcell$10.20$ & $10.44$ & N/A & N/A & 2017-17973 \\
        & \multirow{-2}{*}{T (h)} & $\pm 1.47$ & \gcell$\pm 5.13$ & $\pm 6.68$ & & & \\
      \bugdesc{Use-after-free} \\
      \rowcolor{rowgray} & \# & $2$  & \ycell$4$  & $3$ & $0$ & $0$ & \\
      \rowcolor{rowgray} C & & $12.60$ & \gcell$6.16$ & $11.57$ & N/A & N/A & N/A \\
        \rowcolor{rowgray} & \multirow{-2}{*}{T (h)} & $\pm 5.70$ & \gcell$\pm 5.16$ & $\pm 5.85$ & & & \\
      \rowcolor{rowgray} \bugdesc{Heap buffer overread} \\
      & \# & $1$  & $2$  & $0$ & $0$ & \ycell$11$ & \\
        D & & \gcell$1.77$ & $4.62$ & N/A & N/A & $2.65$ & 2018-5784 \\
        & \multirow{-2}{*}{T (h)} & & $\pm 4.95$ & & & $\pm 4.60$ & \\
      \bugdesc{Uncontrolled memory consumption} \\
      \midrule
    \multicolumn{8}{l}{\textbf{SoX (WAV)}} \\
      \rowcolor{rowgray} & \# & $6$  & $6$  & $4$ & $0$ & \ycell $10$ & \\
      \rowcolor{rowgray} A & & $13.64$ & $11.52$ & $10.90$ & N/A & \gcell$6.88$ & 2019-8355 \\
      \rowcolor{rowgray} & \multirow{-2}{*}{T (h)} & $\pm 2.55$ & $\pm 4.95$ & $\pm 5.76$ & & \gcell$\pm 6.67$ & \\
      \rowcolor{rowgray} \bugdesc{Integer overflow causes improper heap allocation} \\
      & \# & $5$  & $5$  & $4$ & $0$ & \ycell$9$ & \\
      B & & $14.53$ & $10.68$ & $12.25$ & N/A & \gcell$7.15$ & 2019-8357 \\
      & \multirow{-2}{*}{T (h)} & $\pm 1.89$ & $\pm 4.70$ & $\pm 4.67$ & & \gcell$\pm 5.77$ & \\
      \bugdesc{Integer overflow causes failed memory allocation} \\
      \rowcolor{rowgray} & \# & $2$  & $2$  & $1$ & $0$ & \ycell$4$ & \\
      \rowcolor{rowgray} C & & $15.36$ & $15.09$ & $1.61$ & N/A & \gcell$4.64$ & 2019-8354 \\
      \rowcolor{rowgray} & \multirow{-2}{*}{T (h)} & $\pm 2.72$ & $\pm 1.98$ & & & \gcell$\pm 3.23$ & \\
      \rowcolor{rowgray} \bugdesc{Integer overflow causes improper heap allocation} \\
      & \# & $0$  & $0$  & $1$ & $0$ & \ycell$4$ & \\
      D & & N/A & N/A & $12.38$ & N/A & \gcell$8.06$ & 2019-8356 \\
      & \multirow{-2}{*}{T (h)} & & & & & \gcell$\pm 7.04$ & \\
      \bugdesc{Stack buffer bounds violation} \\
      \rowcolor{rowgray} & \# & $30$ & $30$ & $30$ & $30$ & $28$ & \\
      \rowcolor{rowgray} E & & \gcell$0.004$ & $0.005$ & $0.01$ & $0.01$ & $0.57$ & 2017-11332 \\
      \rowcolor{rowgray} & \multirow{-2}{*}{T (h)} & \gcell$\pm 0.001$ & $\pm 0.001$ & $\pm 0.0005$ & $\pm 0.01$ & $\pm 0.53$ & \\
      \rowcolor{rowgray} \bugdesc{Divide-by-zero} \\
    \bottomrule
  \end{tabular}
  \end{adjustbox}
  \end{subtable}
  \begin{subtable}[t]{0.49\textwidth}

  \centering
    \begin{adjustbox}{width=0.95\textwidth}
    \begin{tabular}[t]{p{0.5em}p{3em} rrrrr l}
    \toprule
    \rotatebox{90}{Bug ID} & \rotatebox{90}{Metric} & \rotatebox{90}{CMIN} & \rotatebox{90}{ML-S} & \rotatebox{90}{MS-U} & \rotatebox{90}{Full} & \rotatebox{90}{Empty} & CVE \\
    \midrule
    \multicolumn{8}{l}{\textbf{Poppler}} \\
      \rowcolor{rowgray} & \# & \ycell$13$ & $8$ & $7$ & $0$ & $0$ & \\
      \rowcolor{rowgray} A & & $8.83$ & $8.41$ & \gcell$4.15$ & N/A & N/A & 2019-12293 \\
        \rowcolor{rowgray} & \multirow{-2}{*}{T (h)} & $\pm 5.37$ & $\pm 6.16$ & \gcell$\pm 4.52$ & & & \\
      \rowcolor{rowgray} \bugdesc{Heap buffer overread} \\
      & \# & \ycell$2$ & $1$ & $0$ & $0$ & $0$ & \\
      B & & $13.30$ & \gcell$1.98$ & N/A & N/A & N/A & 2018-21009 \\
        & \multirow{-2}{*}{T (h)} & $\pm 4.57$ & & & & & \\
      \bugdesc{Uncontrolled memory consumption}  \\
    \midrule
    \multicolumn{8}{l}{\textbf{FreeType}} \\
      \rowcolor{rowgray} & \# & \ycell$30$ & $28$ & $25$ & $29$ & $0$ & \\
        \rowcolor{rowgray} A & & $3.69$ & $5.60$ & $7.11$ & \gcell$2.94$ & N/A & 2014-9663 \\
        \rowcolor{rowgray} & \multirow{-2}{*}{T (h)} & $\pm 3.39$ & $\pm 4.40$ & $\pm 4.96$ & \gcell$\pm 2.43$ & & \\
      \rowcolor{rowgray} \bugdesc{Heap buffer overread} \\
      & \# & $30$ & $26$ & $30$ & $28$ & $0$ & \\
      B & & $6.19$ & $5.09$ & \gcell$3.51$ & $5.53$ & N/A & 2015-9290 \\
        & \multirow{-2}{*}{T (h)} & $\pm 3.49$ & $\pm 1.55$ & \gcell$\pm 0.69$ & $\pm 2.54$ & & \\
      \bugdesc{Heap buffer overread} \\
      \rowcolor{rowgray} & \# & $27$ & $27$ & $25$ & \ycell$28$ & $0$ & \\
        \rowcolor{rowgray} C & & $5.37$ & $4.48$ & \gcell$3.49$ & $7.40$ & N/A & 2014-9658 \\
        \rowcolor{rowgray} & \multirow{-2}{*}{T (h)} & $\pm 4.84$ & $\pm 5.50$ & \gcell$\pm 3.76$ & $\pm 4.82$ & & \\
      \rowcolor{rowgray} \bugdesc{Heap buffer overread} \\
      & \# & $15$ & $14$ & $15$ & \ycell$16$ & $0$ & \\
      D & & $9.17$ & $7.26$ & $7.21$ & \gcell$6.85$ & N/A & 2014-9669 \\
        & \multirow{-2}{*}{T (h)} & $\pm 4.02$ & $\pm 4.67$ & $\pm 4.87$ & \gcell$\pm 4.04$ & & \\
      \bugdesc{Integer overflow resulting in invalid length check} \\
      \rowcolor{rowgray} & \# & $13$ & $14$ & \ycell$20$ & $13$ & $12$ & \\
      \rowcolor{rowgray} E & & $9.39$ & $7.46$ & $6.70$ & $7.71$ & \gcell$3.95$ & 2014-2240 \\
        \rowcolor{rowgray} & \multirow{-2}{*}{T (h)} & $\pm 3.02$ & $\pm 3.26$ & $\pm 2.47$ & $\pm 3.62$ & \gcell$\pm 2.17$ & \\
      \rowcolor{rowgray} \bugdesc{Stack buffer overflow} \\
      & \# & $7$  & $6$  & \ycell$13$ & $2$ & $0$ & \\
        F & & $5.75$ & \gcell$4.78$ & $7.22$ & $9.26$ & N/A & 2014-9659 \\
        & \multirow{-2}{*}{T (h)} & $\pm 2.88$ & \gcell$\pm 2.88$ & $\pm 4.43$ & $\pm 10.05$ & & \\
      \bugdesc{Stack buffer overflow} \\
      \rowcolor{rowgray} & \# & $1$  & $0$  & $1$ & $0$ & $0$ & \\
      \rowcolor{rowgray} \multirow{-2}{*}{G} & & $11.26$ & N/A & \gcell$7.57$ & N/A & N/A & \multirow{-2}{*}{2015-9383} \\
      \rowcolor{rowgray} \bugdesc{Heap buffer overread} \\
      & \# & $1$  & $1$  & \ycell$2$ & $1$ & $0$ & \\
        \multirow{-2}{*}{H} & & \gcell$4.29$ & $7.46$ & $6.43$ & $6.32$ & N/A & \multirow{-2}{*}{2015-9381} \\
        & \multirow{-2}{*}{T (h)} & & & $\pm 1.77$ & & & \\
      \bugdesc{Heap buffer overread} \\
    \midrule
    \multicolumn{8}{l}{\textbf{SoX (MP3)}} \\
      \rowcolor{rowgray} & \# & $30$ & $30$ & $30$ & $30$ & $30$ & \\
      \rowcolor{rowgray} A & & $0.06$ & $0.001$ & $0.09$ & $1.21$ & $0.001$ & \multirow{-2}{*}{N/A} \\
        \rowcolor{rowgray} & \multirow{-2}{*}{T (h)} & $\pm 0.01$ & $\pm 0.0001$ & $\pm 0.02$ & $\pm 3.55$ & $\pm 0.0001$ & \\
      \rowcolor{rowgray} \bugdesc{Overlapping source and destination addresses} \\
      & \# & $1$  & $3$  & $2$ & $0$ & \ycell$12$ & \\
      B & & $17.84$ & $5.75$ & $14.20$ & N/A & \gcell$5.47$ & 2019-8355 \\
        & \multirow{-2}{*}{T (h)} & & $\pm 1.29$ & $\pm 3.07$ & & \gcell$\pm 4.61$ & \\
      \bugdesc{Integer overflow causes improper heap allocation}  \\
      \rowcolor{rowgray} & \# & $2$  & $4$  & \ycell$6$ & $0$ & $3$ & \\
      \rowcolor{rowgray} C & & $13.37$ & $9.43$ & $10.13$ & N/A & \gcell$2.54$ & 2017-8373 \\
      \rowcolor{rowgray} & \multirow{-2}{*}{T (h)} & $\pm 5.45$ & $\pm 5.93$ & $\pm 4.75$ & & \gcell$\pm 2.77$ & \\
      \rowcolor{rowgray} \bugdesc{Heap buffer overflow in a 3rd party library} \\
      & \# & $0$  & $3$  & $3$ & $0$ & \ycell$10$ & \\
      D & & N/A & $6.34$ & $15.01$ & N/A & \gcell$6.09$ & 2019-8357 \\
      & \multirow{-2}{*}{T (h)} & & $\pm 0.53$ & $\pm 2.95$ & & \gcell$\pm 4.76$ & \\
      \bugdesc{Integer overflow causes failed memory allocation} \\
      \rowcolor{rowgray} & \# & $0$  & \ycell$3$  & $2$ & $0$ & $2$ & \\
      \rowcolor{rowgray} E & & N/A & \gcell$8.80$ & $13.88$ & N/A & $10.47$ & 2019-8354 \\
      \rowcolor{rowgray} & \multirow{-2}{*}{T (h)} & & \gcell$\pm 4.00$ & $\pm 2.68$ & & $\pm 1.55$ & \\
      \rowcolor{rowgray} \bugdesc{Integer overflow causes improper heap allocation} \\
      & \# & $0$  & $3$  & $2$ & $0$ & \ycell$7$ & \\
      F & & N/A & $8.66$ & $16.05$ & N/A & \gcell$6.68$ & 2019-8356 \\
      & \multirow{-2}{*}{T (h)} & & $\pm 3.84$ & $\pm 0.61$ & & \gcell$\pm 4.02$ & \\
      \bugdesc{Stack buffer bounds violation} \\
      \rowcolor{rowgray} & \# & $3$  & $4$  & $0$ & $0$ & \ycell$25$ & \\
      \rowcolor{rowgray} G & & $17.20$ & $6.57$ & N/A & N/A & \gcell$4.41$ & 2017-18189 \\
      \rowcolor{rowgray} & \multirow{-2}{*}{T (h)} & $\pm 0.12$ & $\pm 6.60$ & & & \gcell$\pm 3.90$ & \\
      \rowcolor{rowgray} \bugdesc{Null pointer dereference} \\
      & \# & $0$  & $0$  & \ycell$3$ & $0$ & $0$ & \\
      H & & N/A & N/A & \gcell$15.01$ & N/A & N/A & 2019-13590 \\
      & \multirow{-2}{*}{T (h)} & & & \gcell$\pm 2.97$ & & & \\
      \bugdesc{Integer overflow causes failed memory allocation} \\
    \bottomrule
  \end{tabular}
  \end{adjustbox}
  \end{subtable}
\end{table*}

\begin{figure*}
  \begin{subfigure}[b]{\textwidth}
    \centering
    \includegraphics[scale=0.5]{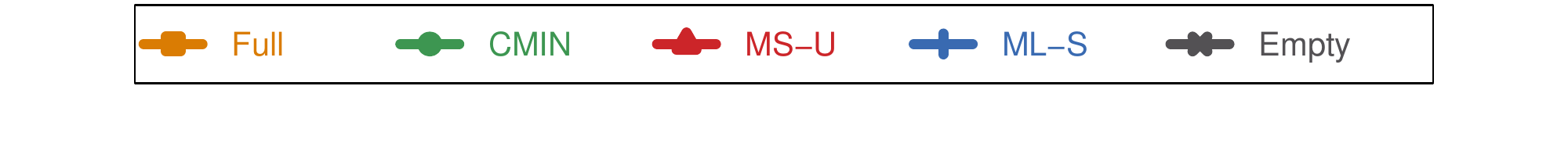}
  \end{subfigure}

  \begin{subfigure}[b]{\textwidth}
    \begin{subfigure}[b]{0.49\textwidth}
      \centering
      \includegraphics[clip, width=\linewidth]{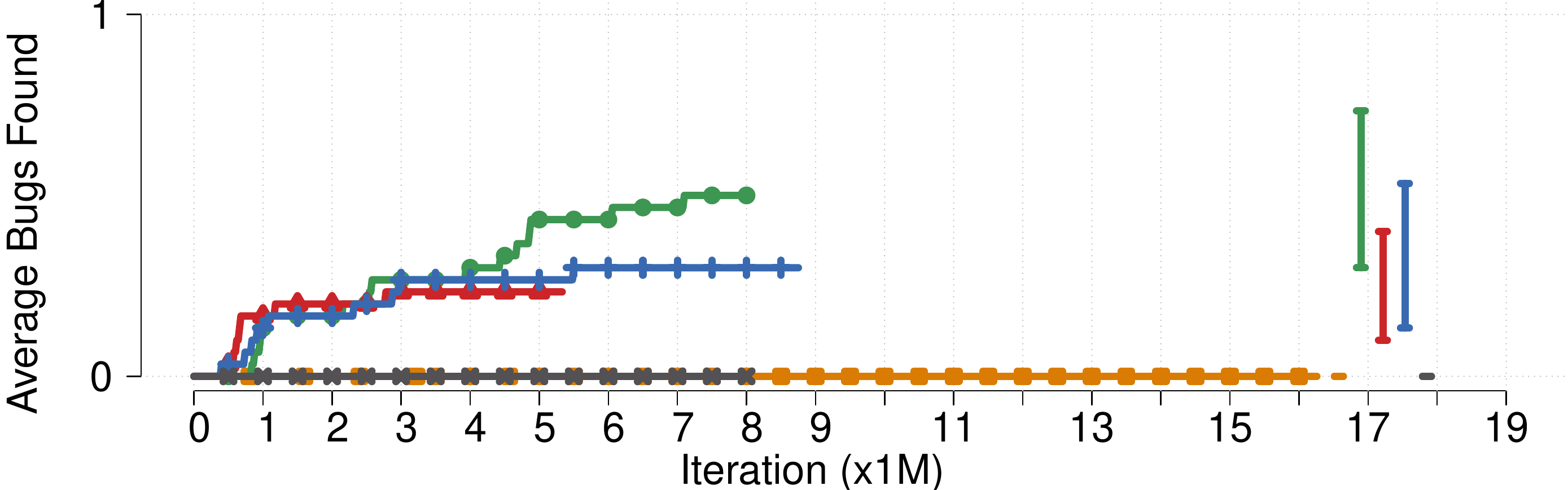}
    \end{subfigure}
    \begin{subfigure}[b]{0.49\textwidth}
      \centering
      \includegraphics[clip, width=\linewidth]{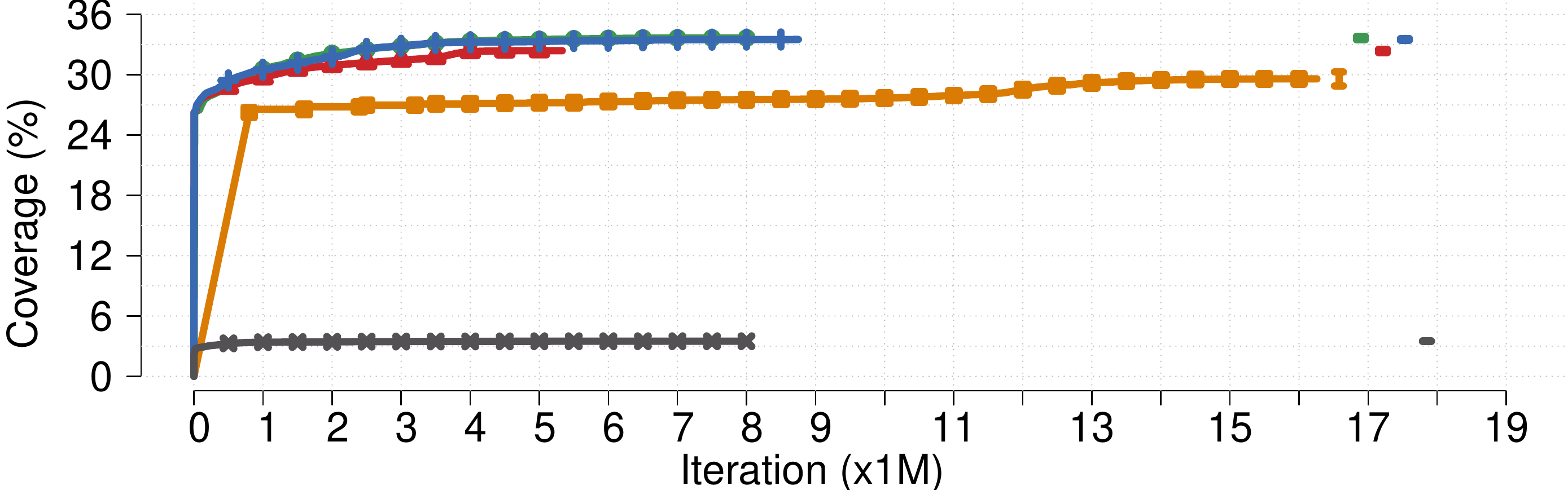}
    \end{subfigure}
    \vspace{-1em}
    \caption{Poppler}
    \label{fig:bug-pdf}
  \end{subfigure}
  \vspace{1ex}
  \begin{subfigure}[b]{\textwidth}
    \begin{subfigure}[b]{0.49\textwidth}
      \centering
      \includegraphics[clip, width=\linewidth]{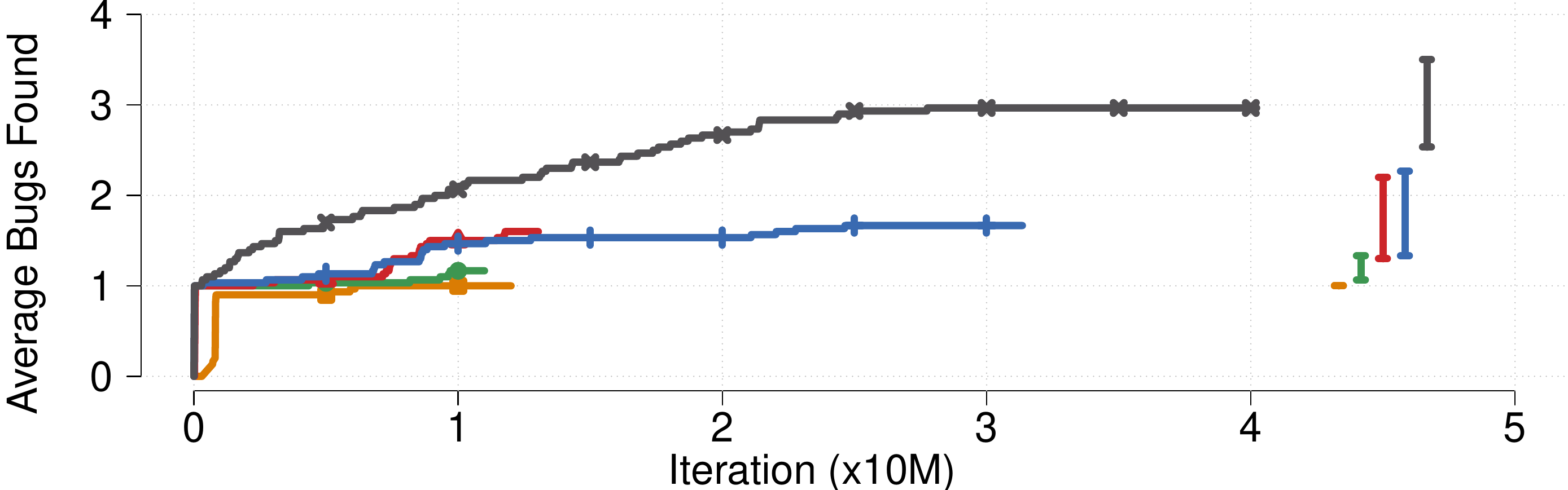}
    \end{subfigure}
    \begin{subfigure}[b]{0.49\textwidth}
      \centering
      \includegraphics[clip, width=\linewidth]{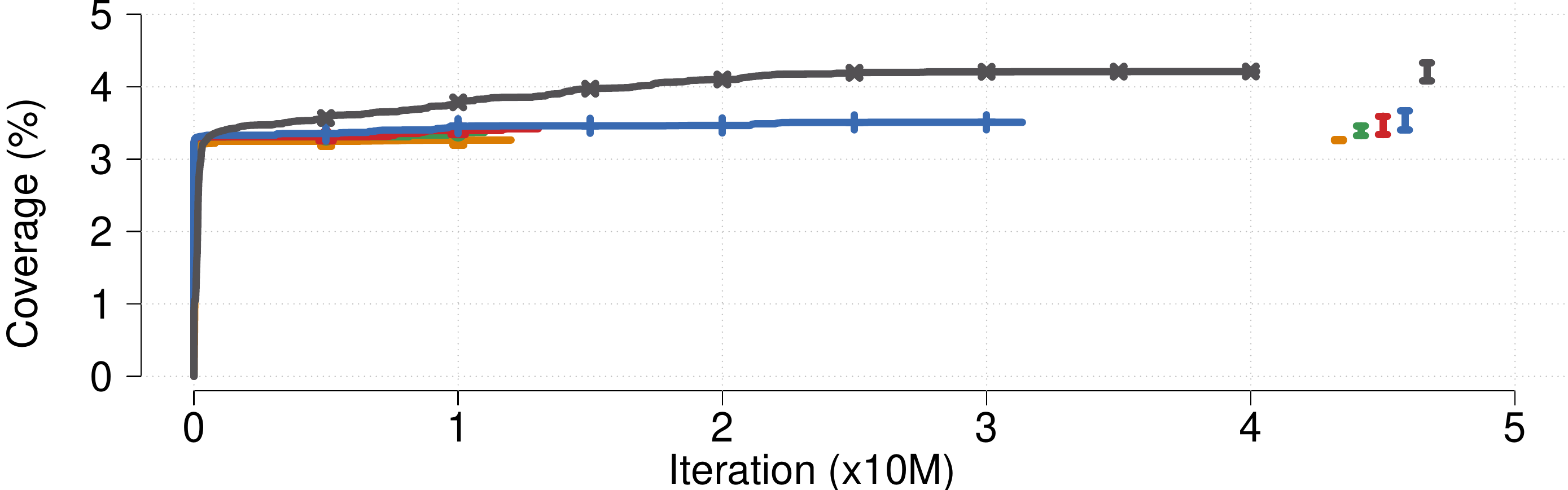}
    \end{subfigure}
    \vspace{-1em}
    \caption{SoX (MP3)}
    \label{fig:bug-sox-mp3}
  \end{subfigure}
  \vspace{1ex}
  \begin{subfigure}[b]{\textwidth}
    \begin{subfigure}[b]{0.49\textwidth}
      \centering
      \includegraphics[clip, width=\linewidth]{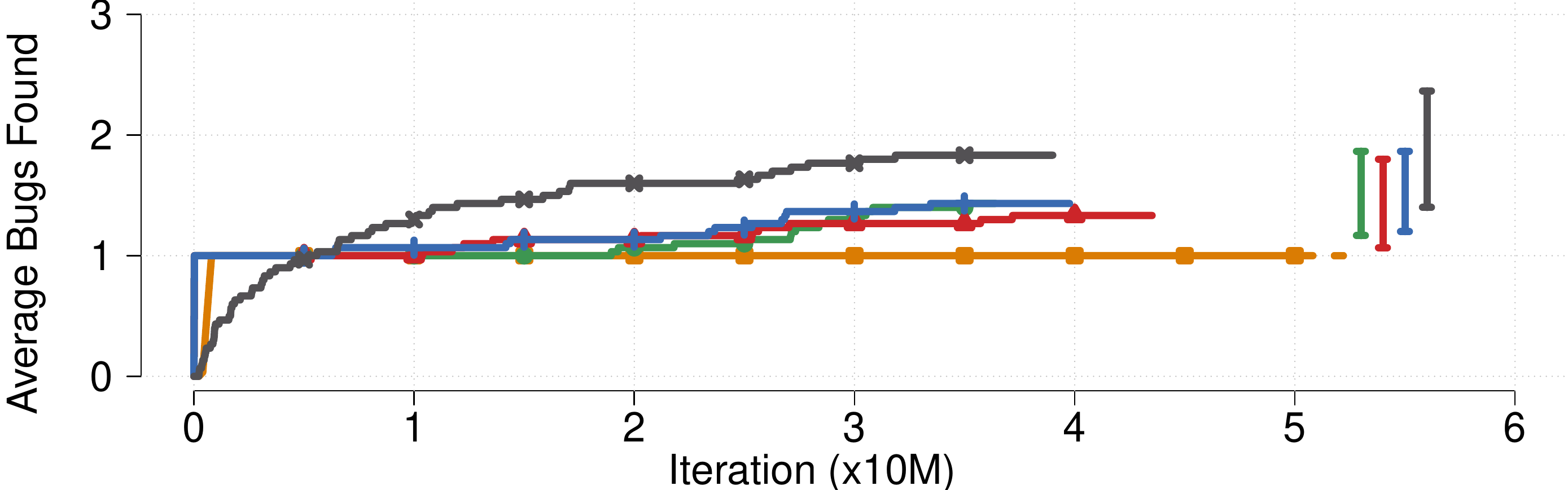}
    \end{subfigure}
    \begin{subfigure}[b]{0.49\textwidth}
      \centering
      \includegraphics[clip, width=\linewidth]{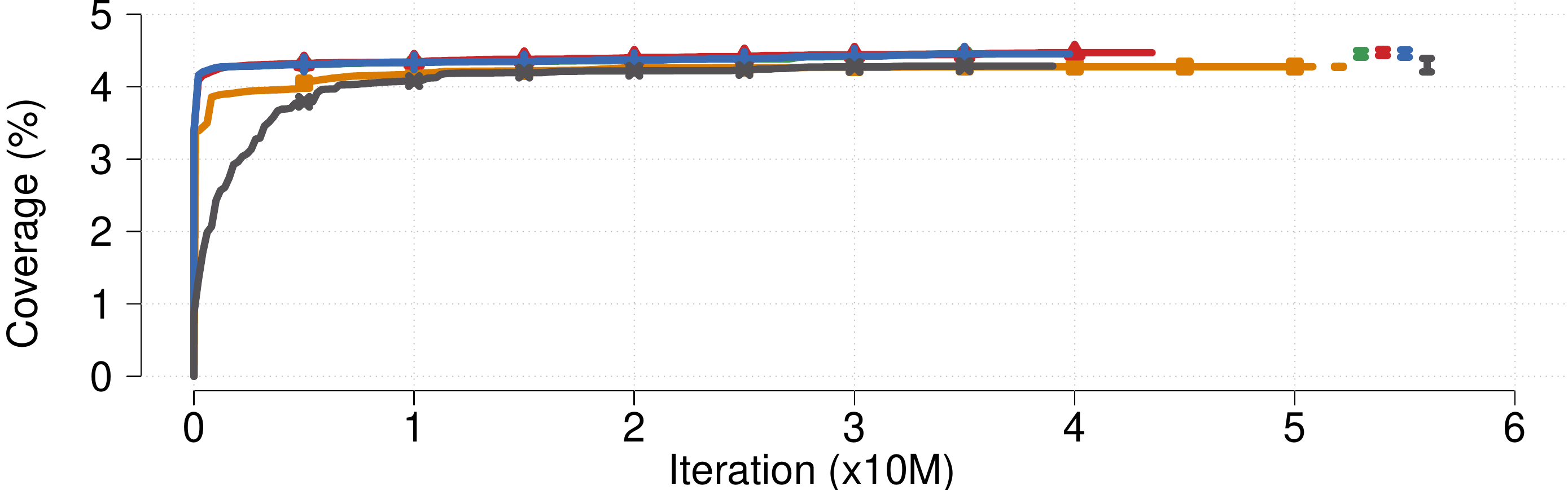}
    \end{subfigure}
    \vspace{-1em}
    \caption{SoX (WAV)}
    \label{fig:bug-sox-wav}
  \end{subfigure}
  \vspace{1ex}
  \begin{subfigure}[b]{\textwidth}
    \begin{subfigure}[b]{0.49\textwidth}
      \centering
      \includegraphics[clip, width=\linewidth]{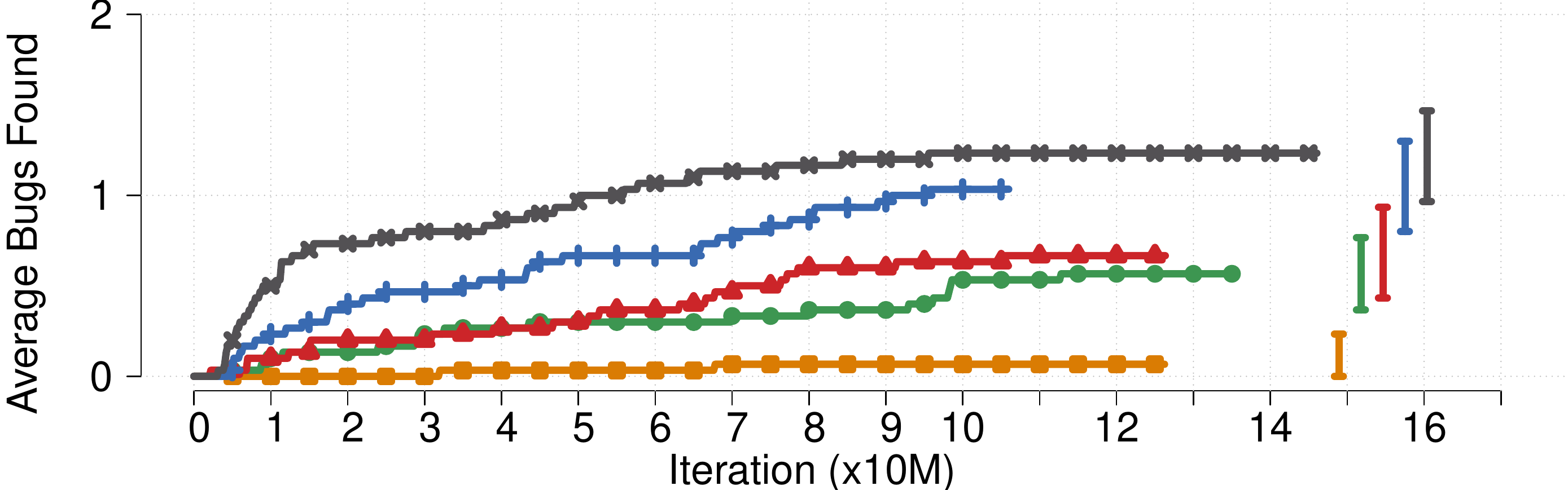}
    \end{subfigure}
    \begin{subfigure}[b]{0.49\textwidth}
      \centering
      \includegraphics[clip, width=\linewidth]{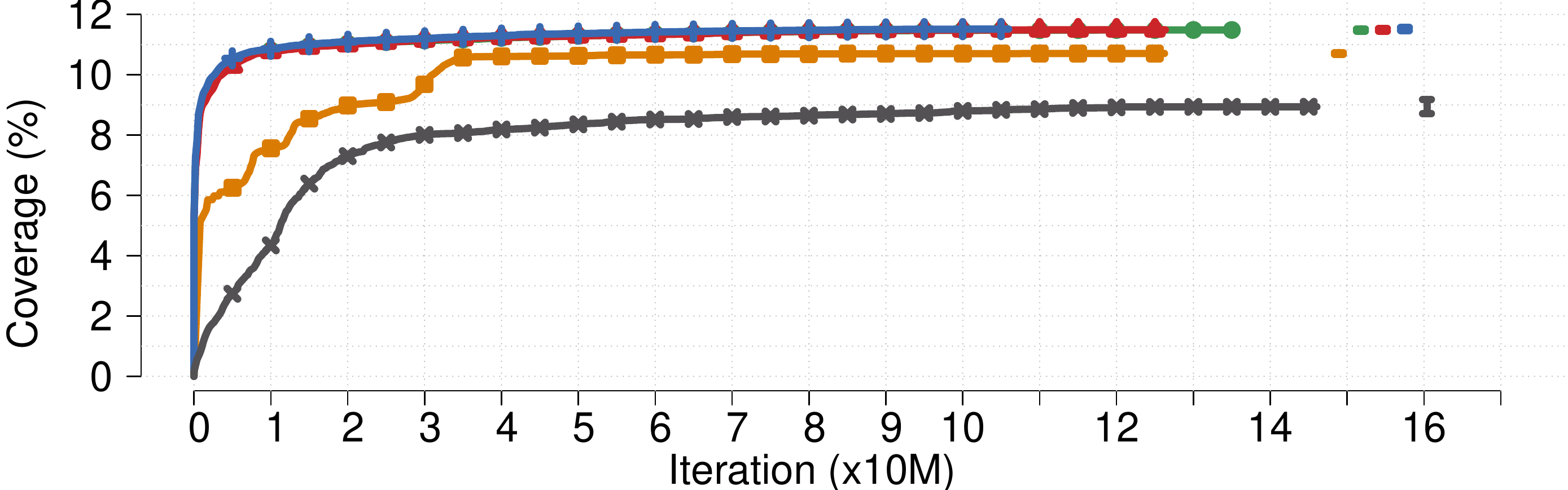}
    \end{subfigure}
    \vspace{-1em}
    \caption{libtiff}
    \label{fig:bug-tiff}
  \end{subfigure}
  \vspace{1ex}
  \begin{subfigure}[b]{\textwidth}
    \begin{subfigure}[b]{0.49\textwidth}
      \centering
      \includegraphics[clip, width=\linewidth]{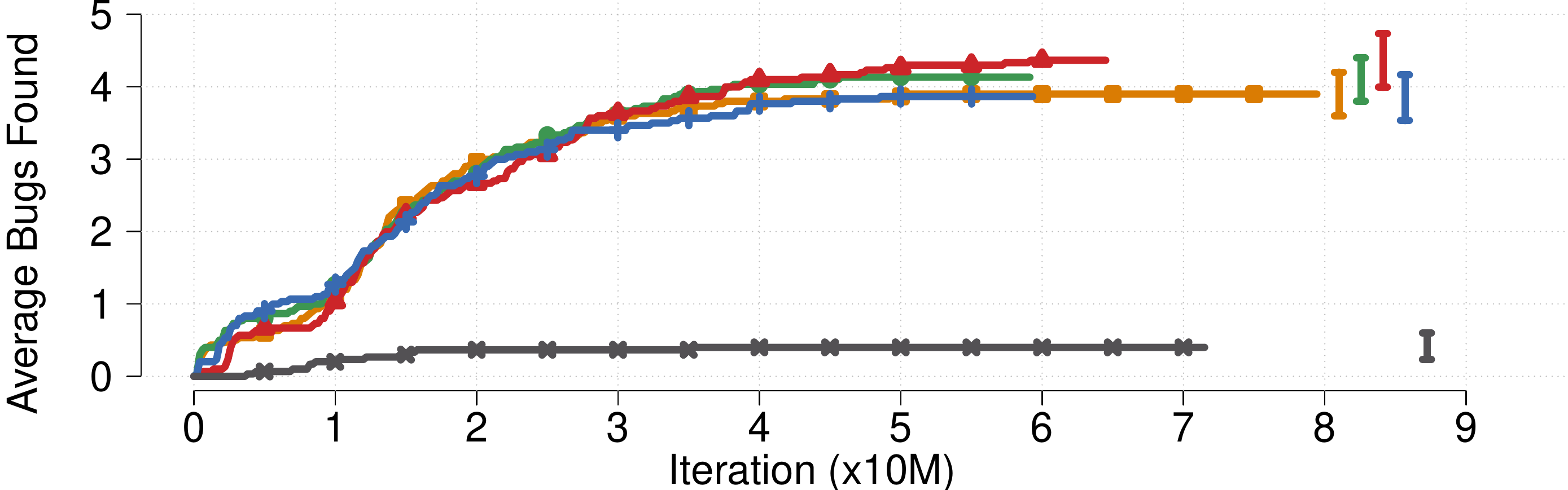}
    \end{subfigure}
    \begin{subfigure}[b]{0.49\textwidth}
      \centering
      \includegraphics[clip, width=\linewidth]{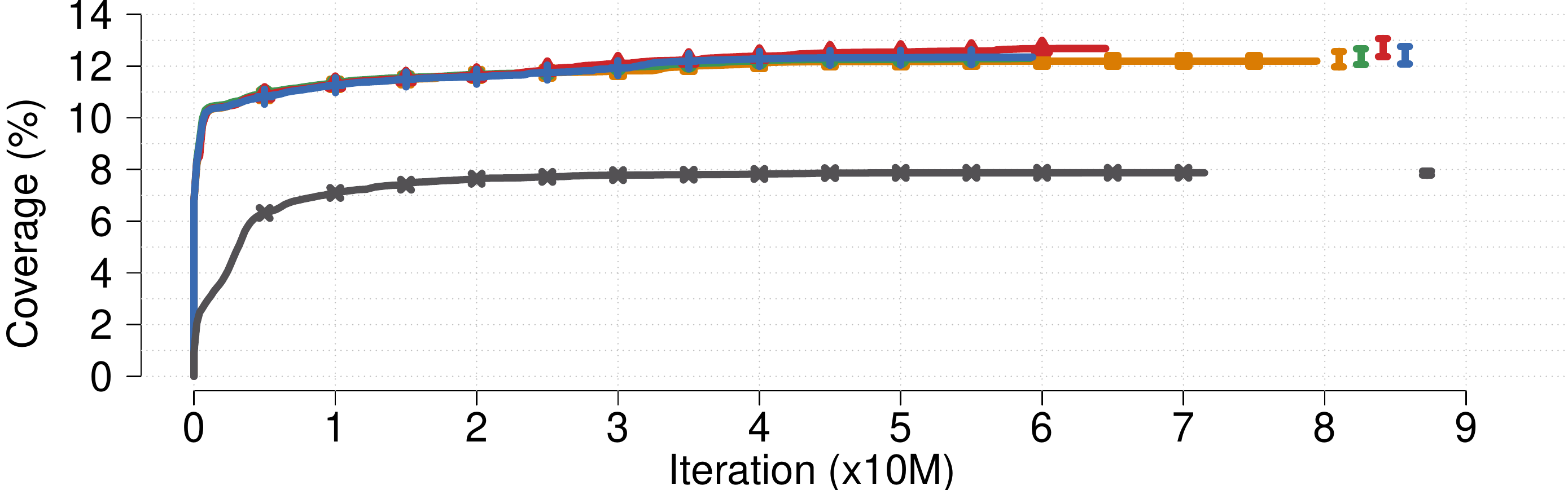}
    \end{subfigure}
    \vspace{-1em}
    \caption{FreeType}
    \label{fig:bug-ttf}
  \end{subfigure}
  \vspace{1ex}
  \begin{subfigure}[b]{\textwidth}
    \begin{subfigure}[b]{0.49\textwidth}
      \centering
      \includegraphics[clip, width=\linewidth]{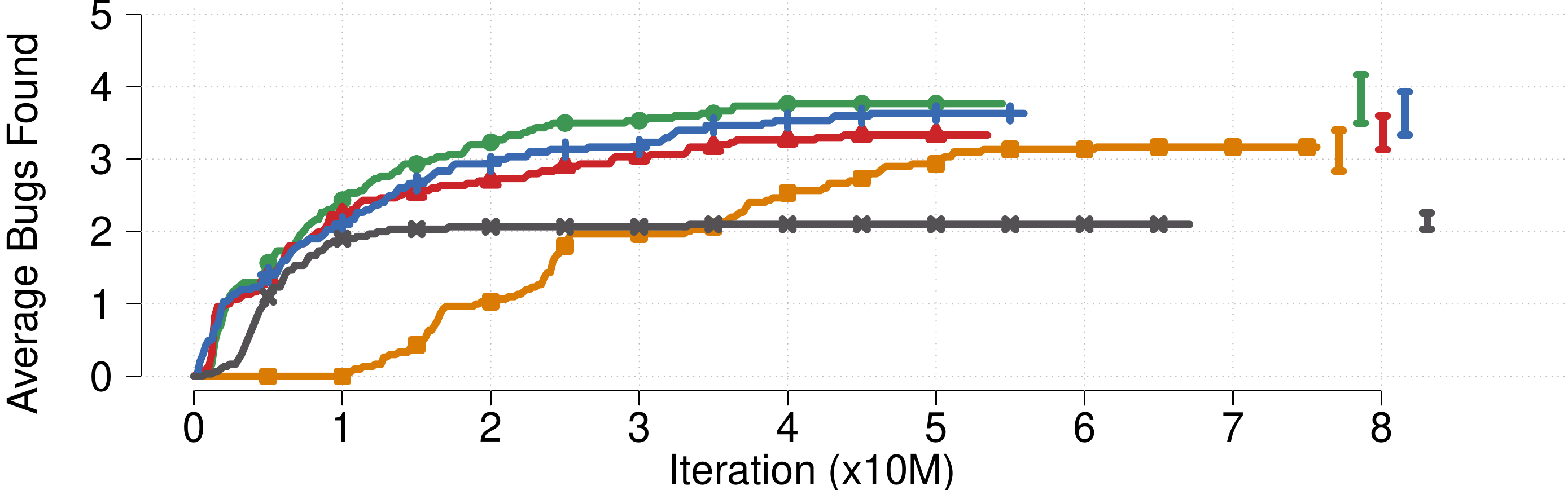}
    \end{subfigure}
    \begin{subfigure}[b]{0.49\textwidth}
      \centering
      \includegraphics[clip, width=\linewidth]{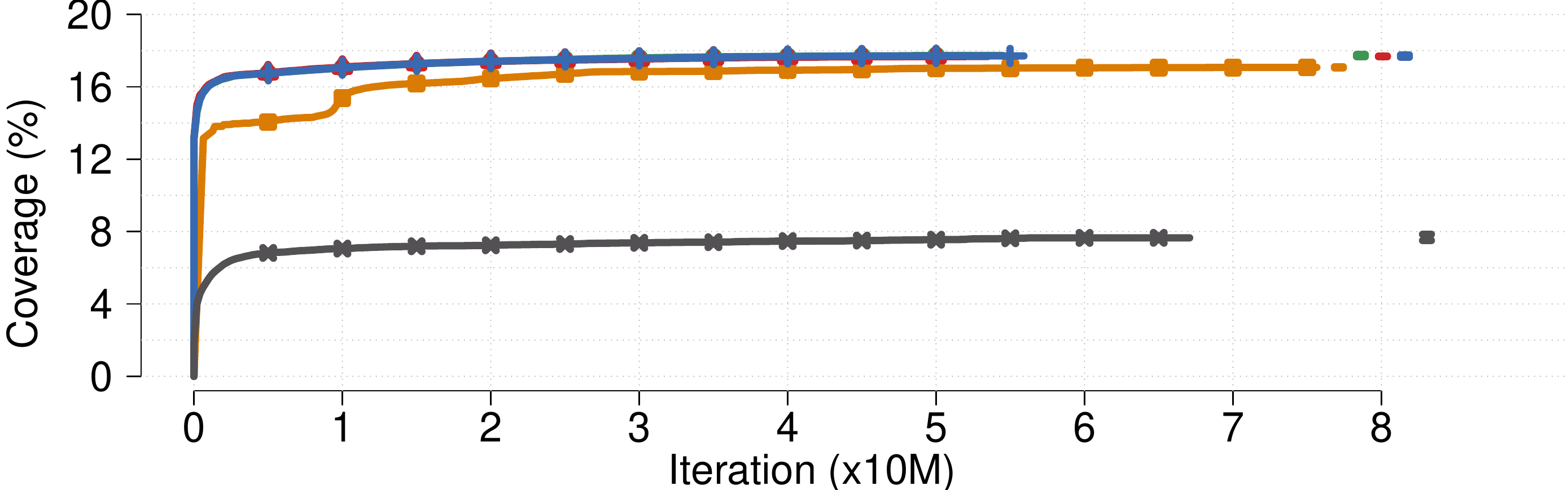}
    \end{subfigure}
    \vspace{-1em}
    \caption{libxml2}
    \label{fig:bug-xml}
  \end{subfigure}
  \caption{Mean number of unique bugs found per trial (left) and code coverage (right) for thirty 18-hour fuzzing trials across the real-world target set.
}
  \label{fig:fuzzing-bugs}
\end{figure*}

The FTS results are largely inconclusive: four of the six coverage
benchmarks are completed without any fuzzing, and many of the bugs are
found consistently by \emph{all} corpora (e.g., libxml2, pcre2, and
re2).  We therefore present the results of fuzzing \numrealprogs
real-world targets, spanning \realnumfiletypes file types (\cref{tab:Targets}).
\Cref{tab:triage-results} contains a summary of all \numbugs bugs that
we found in these targets.  Additionally, \cref{fig:fuzzing-bugs}
shows the \emph{average fuzzer response} for six of the
\realnumfiletypes targets (librsvg produced no bugs, so we omit it).
For each target, there are \numdistills response curves shown (the
mean of thirty trials).  Each curve corresponds to one of the
distillation techniques previously described.

Each plot shows either the cumulative number of unique bugs found or
code coverage against the test iterations.  The intervals displayed on
the right-side of each plot show the \SI{95}{\percent} confidence
intervals.  We use the nonparametric, bias-corrected and accelerated
($\mathrm{BC}_a$) bootstrap interval~\cite{Efron:1987} for these confidence
intervals.  Bug uniqueness is determined by extensive manual triage.

The plots in \cref{fig:fuzzing-bugs} reinforce our choice of trial
length (18 hours) and the number of repetitions (thirty).  Code
coverage has generally reached a steady-state by the time a trial
ends.  This suggests increasing the length of a single trial would
provide little benefit, as AFL has stopped making progress in
exploring these targets.  Conversely, the larger confidence intervals
in the bug yield plots (compared to the confidence intervals in the
coverage plots) illustrates the highly stochastic nature of fuzzing
and emphasizes the need for a large number of repeated trials.  These
plots also show a correlation between the code coverage of a corpus
and its bug yield: higher coverage generally leads to greater bug
yield.  While there are a small number of targets where this is not
true (e.g., libtiff), the differences in bug yield across corpora are
small enough to make this insignificant (e.g., the best performing
libtiff corpora---Empty and ML-S---differ by less than one bug in
their average bug yield).

Once again, when Empty finds a bug, it tends to be the fastest to do
so, while Full remains the slowest at finding bugs.  Empty also has
the highest ``false negative'' rate (as evident from the number of
``N/A'' entries in \cref{tab:triage-results}).  Of the distilled
corpora, ML-S is generally the fastest at finding bugs, while MS-U is
the most reliable.

Before drawing general conclusions, we briefly discuss the
bugs and coverage for each target.  No bugs were found in librsvg, so we
omit it.

\paragraph*{Poppler}
In \cref{fig:bug-pdf}, we see that Full yields twice the number of
executions compared to the other approaches.  We believe this is due
to the fact a very large proportion of seeds in the full corpus are
extremely fast to execute (without necessarily gaining interesting
coverage or bugs).  Such seeds clog the fuzzing queue, leading to low
productivity.  This is reinforced by considering Full's coverage,
which is inferior to all distillation techniques (CMIN, MS-U, and
ML-S).  The remainder of the curves perform similarly both in terms of
mean yield and yield variance.

We found two bugs in this target.  Notably, bug B is never found by
MS-U or Empty.  However, CMIN and ML-S rarely find it---less than
three times each out of thirty trials---indicating that this bug is
generally difficult to discover.

\paragraph*{SoX}

A highlight for both MP3 and WAV file types is the effectiveness of
Empty (particularly its bug-finding speed), despite having a large
process variance (as observed by the confidence intervals).  Nine bugs
were found in this target (across both file types).  Of these nine,
only one was previously reported.

Focusing on the MP3 results, ML-S finds three bugs (D--F) (three times
each) that CMIN does not.  Similarly, bug G is found by all corpora
\emph{except} MS-U and Full.

Interestingly, the MS-U corpus is the only one to find bug H.  We
traced bug H back to its source seed in the corpus (using the parent
seed identifier embedded within the file name of the crashing input)
and found that this particular bug can be attributed to one of two
seeds (i.e., the bug was found by one seed in one trial, and a
different seed in another two trials) that only MS-U selects.  Using
Principle Component Analysis (PCA) on the corpus code coverage we
identified three seeds selected by ML-S that exhibit similar behavior
(i.e., achieve similar code coverage) to the two seeds that find bug
H.  Intuitively, one might expect that these three seeds would also
lead to bug H.  However, compared to the two seeds that found bug H,
our analysis found that these seeds were rarely scheduled by AFL:
$\sim 2.5$ million iterations (mean over thirty trials), compared to
$\sim 8$ million iterations (mean over thirty trials) for the two
seeds that found the bug---a result determined to be an artifact of
the seed's filename, which impacts the fuzzer's scheduling.
This explains why an ML-S-distilled corpus is unable to find bug H.

The WAV bugs mostly intersect with the MP3 bugs.  However, the WAV
file type also uncovers an additional divide-by-zero error.  Bug C is
not triggered when fuzzing WAV files because the external library
(libmad) is not used by the WAV codec.

\paragraph*{libtiff}
Similar to SoX, Empty also performs surprisingly well on this target.
This is closely followed by ML-S.  Interestingly, bug A---found by
both CMIN and MS-U in less than half of the trials, but by ML-S
in~\SI{70}{\percent} of trials---is only evident because we target~32-bit
\texttt{x86}.  The libtiff maintainers report that the undefined
behavior at the root of this bug does not manifest on~64-bit targets.

Bug D---an uncontrolled resource consumption, caused by an
infinite loop in the image file directory linked list---is discovered
rarely by CMIN and ML-S (less than~\SI{7}{\percent} of trials), and
never by MS-U and Full.  In contrast, this bug is found more
frequently by Empty (\SI{37}{\percent} of trials).  Notably, the
initial Empty file does not contain any image file directories, while
all of the TIFF files in our distilled corpora do.  We hypothesize
that AFL's mutations break existing directory structures (leading to
parser failures), whereas Empty is able to construct a (malformed)
directory list from scratch.  These mutations eventually lead to a
loop in the list, causing the uncontrolled resource consumption.

\paragraph*{FreeType}
Unlike the other targets, FreeType's Full corpus is competitive.
In particular, the full corpus only contains~\num{466} seeds---i.e., it is
relatively small.  This suggests that distillation is only worthwhile
when there are many seeds in the corpus.

The bug yield is relatively consistent across the various corpora and
no single distillation technique is a clear winner, although Empty is
clearly inferior, plateauing early with low coverage.  Once again, for
the single bug that Empty does find (bug E), it finds it the fastest.
We targeted an older version of FreeType (v2.5.3 from~2014), and all
discovered bugs have since been fixed.

\paragraph*{libxml2}
Despite yielding a relatively high number of iterations, Empty
performs the worst in terms of discovering bugs and maximizing
coverage.  This is similar to FreeType, suggesting that the empty
corpus performs poorly on structured data.

Similar to FreeType, we targeted an older version of libxml2 (v2.9.0
from~2012), and all discovered bugs have since been fixed.

\subsection{Summary of Results}\label{sec:findings}

\textbf{CMIN} produces \emph{significantly} larger corpora compared to
ML-S and MS-U.  It also had the highest false negative count of the
distilled corpora (it failed to find seven of the bugs in the
real-world target set, compared to ML-S and MS-U, which failed to find
five and six bugs respectively).  However, CMIN does outperform ML-S
on 11 of the \numbugs bugs in \cref{tab:triage-results} by finding
them more reliably.  This bug-finding reliability is important due to
the highly-stochastic nature of fuzzing.

\textbf{ML-S} outperforms CMIN and other approaches overall, in terms
of mean bug-finding speed. It out-performed CMIN on~24 of the bugs in
\cref{tab:triage-results}, and MS-U on~22
bugs. 
Notably, ML-S found five bugs that were never found by
MS-U. 
This bug-finding speed is important when a fuzzing campaign is limited
in the time that it can run for.

\textbf{MS-U} corpora have good performance, in general: they were the
fastest at finding eight of the bugs from \emph{both} benchmark
suites, and found four bugs that were never found by
ML-S. 
Both MS-U and ML-S have similar measures of bug-finding reliability.

\textbf{Full} is recommended only when the total number of seeds is
small---i.e., on the order of a few hundred or less.  When there are
thousands of seeds in the collection corpus it is imperative that some
form of distillation is applied.  This is particularly evident in the
FTS' guetzli target: AFL never completed the initial run of all~\num{120000}
seeds in the corpus.  These results agree with those
found by \citet{Rebert:2014}.

\textbf{Empty} performs surprisingly well on average. However,
individual trials may differ wildly from the mean.  It performed best
on highly unstructured data (e.g., audio codecs) and poorly on
structured data (e.g., PDF).  It may make sense to always add the
empty seed to any fuzzing corpus and rely on the fuzzer's own
reinforcement learning to decide if the empty seed is valuable or not.
Alternatively, giving the empty seed its own, separate, campaign and
forcing the fuzzer to attack the one seed's descendants may be what it
is required to see the speedy results.  Certainly, in the case where
the empty corpus finds a particular bug, it tends to be the fastest to
find it.

\paragraph{Constructing the Empty Seed}

Another important consideration (perhaps counter-intuitively) is
\emph{what} the empty seed contains.  We found that fuzzing with a
purely empty file led to very poor results (this was most evident when
we first fuzzed SoX), as AFL was not able to overcome many of the
parser's format checks.  This led us to construct minimal seeds
(examples of which are given in \cref{app:EmptySeeds}) that passed
these initial format checks but did not contain actual data that could
be corrupted by random mutation, potentially breaking these same
format checks.  We hypothesize that this minimal seed is what leads
Empty to find libtiff's bug D with a relatively high level of
reliability.


\section{Conclusions}\label{sec:Conclusion}

Our premise is that the choice of fuzzing corpus is a critical
decision made before a fuzzing campaign begins.  Our results provides
ample confirmation that this is indeed the case, and demonstrate that
coverage-based distillation techniques such as \tool and \minset yield
superior outcomes.

We have performed extensive experiments (over \numcpuyears worth) to
produce findings that provide statistically reliable support for our
claims.  On the basis of theoretical reasoning about mutation-based
fuzzing, we developed a new algorithm for solving the corpus
distillation problem.  We further predicted that distillation using
file size weighting would significantly reduce the mutation search
space and result in more effective fuzzing.  This was shown to be the
case.  Our comparison of \numdistills corpus distillation techniques
shows that no single technique produces all of the bugs that we found.
\tool and \minset appear to have their own strengths, and both
generally outperform \cmin.  Our open-source tools \tool and \bvtool
are freely available along with our collection corpus trace
data. These are available at URLs \bitlytools and \bitlydata.  We look
forward to others experimenting and building on these techniques.

We add to the knowledge of how to perform effective fuzzing in
practice:
\begin{itemize}[nosep]
\item Maximizing fuzzing yield is achieved by using \tool weighted by
  file size or \uminset.
\item Compared to \uminset and \cmin, \tool weighted by file size is
  (on average) the fastest at finding bugs.
\item Less utility is provided by \cmin: it produces the largest
  corpora, finds fewer bugs, and is (on average) the slowest at
  finding bugs.
\item Campaigns should avoid fuzzing with a large collection
  corpus---i.e., on the order of a thousand files or more.
  Conversely, if the collection corpus is small, then distillation is
  not helpful.
\end{itemize}
We also triaged the crashes from the real-world targets, finding
\numbugs bugs, \numnewbugs that are new.  We have reported all
\numnewbugs new bugs and received CVEs for \numcves of them.

\subsection{Future work}

Some of our results raise new questions in response to observed
unexpected behaviors.  For example, the performance of the empty
corpus shows unexpected volatility.  Depending on the target, the
approach can show outstanding performance or the opposite.  The
reasons are unclear and require further investigation. 
However, since fuzzers invariably reward performing seeds, it makes
sense for practitioners to include the empty seed in their fuzzing corpus
and rely on the fuzzer to adapt.


\bibliographystyle{ACM-Reference-Format}
\bibliography{moonlight}

\appendix

\section{The ``Empty Seed'' Corpus}\label{app:EmptySeeds}

As discussed in \cref{sec:experiment}, we use a small, hand-constructed input when fuzzing the ``empty seed''.
The TTF, XML, regex (re2 and pcre2 targets), MP3, and JSON empty seeds contain a single line-break character (``\verb|\n|'').
For the remaining filetypes, the empty seeds are described below.

\medskip
\noindent
The empty SVG:
\begin{quote}
\begin{verbatim}
<svg></svg>
\end{verbatim}
\end{quote}
Similarly, the empty PDF:
\begin{quote}
\begin{verbatim}
%PDF-1.7
1 0 obj
<< /Type  /Catalog
>>
endobj
trailer
<<
/Root 1 0 R
>>
%%EOF
\end{verbatim}
\end{quote}
The empty TIFF contains only the byte-order identifier---it does not contain any image file directories:
\begin{quote}
\begin{verbatim}
II
\end{verbatim}
\end{quote}
In contrast to those above, WAV files do not have a textual representation, hence we use a
combination of ASCII and hexadecimal values (using Python string notation) to
illustrate the empty WAV seed (line breaks have been added for clarity):
\begin{quote}
\begin{verbatim}
RIFF\x24\x00\x00\x00
WAVEfmt \x00\x00\x00\x00
data\x00\x00\x00\x00
\end{verbatim}
\end{quote}
The empty GZIP is an archive containing an empty file.

Finally, the empty JPEG, PNG, and OGG were obtained from the following websites (respectively):

\begin{itemize}[noitemsep]
  \item \url{https://stackoverflow.com/a/30290754}
  \item \url{https://garethrees.org/2007/11/14/pngcrush/}
  \item \url{https://commons.wikimedia.org/wiki/File:En-us-minimal.ogg}
\end{itemize}


\end{document}